\documentclass[aps,twocolumn,showpacs,floatfix,amssymb]{revtex4}
\usepackage{graphicx}
\begin{document}

\title{Electronic spectrum and  superconductivity  in the  extended $t$--$J$--$V$
model}
\author{Nguen Dan Tung$^{a,b}$, A.A. Vladimirov$^{a}$, and  N. M. Plakida$^{a}$ }
 \affiliation{ $^a$Joint Institute for Nuclear Research,
141980 Dubna, Russia}
 \affiliation{$^{b}$
{ Institute of Physics,
Viet Nam Academy of Science and Technology,
10 Dao Tan, Ba Dinh, Hanoi 1000, Viet Nam} }

\date{\today}

\begin{abstract}
A consistent microscopic theory of superconductivity for strongly correlated
electronic systems is presented within the extended $t$--$J$--$V$ model where the intersite Coulomb repulsion  and the electron-phonon interaction are taken into account.  The exact Dyson equation for the normal and anomalous (pair) Green functions is derived for the projected (Hubbard) electronic operators. The equation  is solved in the self-consistent Born approximation  for the self-energy.  We obtain the $d$-wave pairing with high-$T_c$ induced by the strong kinematical interaction of the order of the kinetic energy $\sim t$ of electrons with spin fluctuations which is much larger
than the exchange interaction $J$.  The  Coulomb repulsion  and the electron-phonon interaction  give small contributions  for the $d$-wave pairing. These results  support the spin-fluctuation mechanism of high-temperature superconductivity in cuprates previously proposed in phenomenological models.
\end{abstract}


\maketitle

\section{Introduction}
\label{sec:1}
Since the discovery of the high-temperature superconductivity (HTSC) in  cuprates by Bednorz and M\"uller~\cite{Bednorz86} many theoretical models were proposed to reveal the mechanism of HTSC but a commonly accepted one is still lacking (see, e.g.~\cite{Schrieffer07,Plakida10}).
The main problem in a theoretical study of the cuprate superconductors is that strong electron correlations  preclude  application of the conventional Fermi-liquid approach in  description of their  electronic structure~\cite{Fulde95}. They are Mott-Hubbard  (more accurately, charge-transfer) antiferromagnetic (AFM) insulators  where the conduction band due to the strong Coulomb interaction splits into two subbands of singly-occupied and doubly-occupied states in the lattice. In this case conventional electronic operators cannot be used and composite or projected electronic operators for subbands   should be introduced. To cope with the problem of unconventional character of the projected electronic operators various methods have been applied  in investigation of strongly-correlated electronic systems  (for a review see~\cite{Avella12}).

The first model of strongly correlated electrons revealing superconductivity is the  $t$--$J$ model proposed by Anderson~\cite{Anderson87}. It can be derived from the Hubbard model~\cite{Hubbard63} in  the strong correlation limit~\cite{Bogoliubov49,Bulaevskii68,Spalek78,Izyumov97}. In the $t$--$J$ model superconductivity occurs at finite doping in the spin-liquid of resonating valence-bond states (RVB) due to the AFM superexchange interaction $\,J$. The RVB scenario was considered later using the gauge theory~\cite{Baskaran87}, the mean-field $t$--$J$ model  with the renormalized hopping $t$ and  exchange interaction $J $ parameters ~\cite{Zhang88}, the variational Monte Carlo method for  Gutzwiller wave functions ~\cite{Paramekanti04,Anderson04}.
In Ref.~\cite{Zhang88a} the $t$--$J$ model was formulated as an effective Hamiltonian for the hole-doped superconducting cuprates.  Since then the low-energy electronic spectrum, superconductivity,  spin excitations in cuprates have been  considered   within the $t$--$J$ model by numerous authors.

Extensive numerical studies  have been performed by various methods, such as  Lanczos diagonalization of small clusters,  quantum Monte Carlo simulations of two-dimensional lattices, cluster approximations  (for reviews see~\cite{Dagotto94,Jaklic99,Bulut02,Maier05,Scalapino07} and references therein).  A delicate balance between
superconductivity and other instabilities, AFM, spin-density wave, charge-density wave,
etc., was found~\cite{Scalapino07}.

To take into account strong Coulomb correlations in the Hubbard model    the dynamical mean field theory (DMFT)  was proposed~\cite{Georges96,Kotliar06,Vollhardt12}. It was used to consider the Mott-Hubbard transition from a conventional metallic state to an insulating state. However, to study phase transitions to AFM state or superconductivity the theory should be generalized to take into account short-range correlations. It  was done within the dynamical cluster approximation (DCA) formulated in the reciprocal space,~\cite{Maier06,Gull09,Gull10} and the cluster DMFT (or cellular DMFT) where the impurity single-site in DMFT is replaced by a finite cluster of lattice sites (see, e.g.,~\cite{Stanescu06,Haule07,Kancharla08,Civelli09,Senechal12}). In the  cluster perturbation theory (CPT) an exact diagonalization of the electronic spectrum in a finite cluster (usually, $ 4\times4$ sites in the 2D model) is found and then coupling between of the clusters is taken into account~\cite{Gros93,Senechal00,Senechal02,Senechal12a,Kuzmin14,Kohno15,Kuzmin20}.
A variational cluster approximation (VCA) was also proposed \cite{Potthoff03,Potthoff03a,Aichhorn07} where it was shown that the CPT and the cellular DMFT are limiting cases of a more general cluster method.  A two-particle self-consistent approach  was developed in Refs.~\cite{Vilk94,Vilk95,Vilk97,Tremblay06,Davoudi07,Tremblay12}.
To take into account short-range correlations  a generalized DMFT approach was proposed in Refs.~\cite{Sadovskii01,Sadovskii05,Kuchinskii05,Kuchinskii06}. In the theory  a momentum dependent self-energy $\Sigma_{\bf k} (\omega)$ was included in addition to the DMFT single-site self-energy $\Sigma (\omega)$. This approach permits to describe a pseudogap formation near the Fermi level of the quasiparticle band.

In  cluster theories, it was possible to describe the electronic spectrum with formation of the pseudogap state and the arc-type Fermi surface at low doping.  The $d$-wave superconductivity was also found but it was difficult to disclose  the mechanism of the pairing. The role of the intersite Coulomb repulsion and the electron-phonon interaction are also difficult to include in the cluster calculations. To cope with problems, analytical approaches   were also considered in several studies using various approximations for strong CI within the Hubbard-type models.

To deal with the projected character of the electronic operators which imposes   local constraints of no double occupancy  of  lattice sites in the $t$--$J$ model,   the slave-boson (-fermion) technique was proposed (see~\cite{Suzumura88,Kotliar88,Grilli90,Arrigoni94,Lee06,Ogata08} and references therein). In the mean-field approximation (MFA), commonly used in this method, the local constraints are approximated by a global one, that reduces the problem to conventional fermions and bosons in the
mean field.
In Ref.~\cite{Feng15} the slave-boson representation was considered beyond the MFA for the extended $t$-$J$ model. A kinetic-energy driven mechanism of superconductivity for the  fermion-spin theory was proposed  where the pairing of fermions is induced  by spin excitations described by  slave bosons. However, as in the conventional slave-boson theory the local constraint of no double occupancy was not treated rigorously.

Several other technique for strongly correlated system were also proposed, as e.g., a continued fraction representation for the GFs  in Ref.~\cite{Sherman02}. It was  used in studies of spin excitations and hole spectrum in Refs.~\cite{Sherman03,Sherman04}. We mention also a  diagram method for the Hubbard model suggested  in Refs.~\cite{Vladimir90}  as a site cumulants expansion in terms of hopping parameters. In this  case a moderate to strong Hubbard repulsion can be considered. Using this technique in Ref.~\cite{Sherman06} the electronic spectrum was found in the one-loop approximation which shows  the four-band structure as observed in numerical calculations. The strong coupling diagram technique was used in Ref.~\cite{Sherman18} for investigating the influence of spin and charge fluctuations on electron spectra of the 2D $t$--$U$ Hubbard model.
 A dual fermion approach was proposed in Refs.~\cite{Rubtsov08,Hafermann09,Rubtsov09}.

A formally rigorous method to treat the unconventional commutation relations
for the projected electronic operators is based on the Hubbard operator (HO) technique~\cite{Hubbard65} (a generelazation of the HO representation for non-canonical degrees of freedom is given in \cite{Quinn20}). The diagram technique for the HOs  was developed  to study the Hubbard and $t$--$J$ models~\cite{Slobodyan74,Zaitsev76,Izyumov89,Ovchinnikov04}. A  superconducting pairing due to the kinematical interaction in the Hubbard model in the limit of strong electron correlations ($U \to \infty $) was first considered in Refs.~\cite{Zaitsev87}. In the lowest order diagrams for the two-particle vertex equation which is equivalent to  the  MFA for the superconducting order parameter gives only the $s$-wave pairing.

A technically simpler method in comparison  with the  diagram technique is the projection technique~\cite{Plakida12} in the equation of motion method  for the Green functions (GFs)~\cite{Zubarev60} based on the Mori memory function approach~\cite{Mori65}. Applying this method in terms of the HOs in Refs.~\cite{Plakida89,Yushankhai91}  the $d$-wave pairing for the $t$--$J$ model was found in the MFA. It was also shown that the $s$-wave pairing is prohibited since it violates the restriction of no double occupancy.  Supercoducting pairing in the  singlet band of the Emery model was considered in Ref.~\cite{Plakida94}.  In Refs.~\cite{Mancini04,Avella07,Avella07a,Avella12a} the equations of motion method for the GFs in the composite operator representation, similar to the HOs, was  used in studies of the Hubbard model in the limit of strong correlations.
 The electronic spectrum, spin excitations and phase transitions were analyzed.

The  MFA in the first order projection technique was considered in many studies of electronic and spin-excitation spectra in more complicated $t$--$J$ models.
In Ref.~\cite{Valkov02} a strong effect  of the three-site interaction $H_3$ in the  $t$--$J^{*}$ model on the  $d$-wave superconductivity was stressed and  a modification of the superconducting order parameter was found in Ref.~\cite{Valkov03}.
In Refs.~\cite{Jedrak10,Jedrak11} the renormalized mean-field theory for the $t$--$J$ model was formulated and comparison with experiments in cuprates was performed. Using the LDA and tight-binding approximation for the La$_{2-x}$Sr$_x$CuO$_{4}$  an effective  $t'$--$t''$--$J^{*}$ model was derived in Ref.~\cite{Ovchinnikov09}. The model was used to study the Lifshitz quantum phase transitions and transformation of the Fermi surface with hole concentration. Influence of the interlayer tunneling  $t_2$ on the electronic structure of the bilayer cuprates with hole concentration and strong magnetic fields was considered in Ref.~\cite{Ovchinnikov11}. Superconducting $T_c$  and spin correlations within $t'$--$t''$--$t_2$--$ J^{*}$ model were found in Ref.~\cite{Makarov12}. An effective model for electron-phonon and spin-phonon interactions for the original $p$--$d$   model~\cite{Gaididei88}  was derived in  Ref.~\cite{Ovchinnikov05}. Considering only holes in the singly-occupied  Hubbard subband the low-energy $t$--$J^{*}$ part of the model with electron-phonon and spin-phonon interactions in terms of the HOs was also proposed. The isotope effect in the $t$--$J^{*}$ model with electron-phonon coupling was discussed in Ref.~\cite{Shneider09}  and in Ref.~\cite{Makarov12} for the bilayer cuprates.
 Superconductivity in the two subband  $t$--$J$ model for the honeycomb lattice was considered recently in many publications. In ref.~\cite{Vladimirov19} the singlet order parameter for the $d+id'$ pairing was obtained and the superconducting $T_c$ as a function of doping was calculated.

In studies of electronic spectrum and superconductivity within the  $t$--$J$ model in MFA the exchange interaction $J$ was considered as the origin of electron coupling to the spin system.
To go beyond the MFA  higher order contributions to  electron interaction with spins should be considered. Applying the projection technique in the equation of motion method for the GFs in terms of the HOs a microscopic theory of spin-fluctuation superconducting pairing was  proposed in Refs.~\cite{Plakida99,Prelovsek05}. The Dyson equation for the normal and anomalous GFs was derived where a new energy scale caused by the kinematical interaction of electrons with  dynamical spin fluctuations were found.  The interaction is determined by the kinetic energy of electrons given by the hopping parameter $t$, much larger then the exchange interaction $J$. Calculation of the normal part of the self-energy operator  brings about the renormalization of the  electronic spectrum where at low doping the pseudogap and  the arc-type Fermi surface appear~\cite{Plakida99,Prelovsek97,Prelovsek01}.  Suppression of the quasiparticle weight in the equation for the superconducting gap results in lowering of the superconducting transition temperature $T_c$ in comparison with the MFA results.
Similar spin-fluctuation superconducting pairing was  proposed in Ref.~\cite{Onoda01}. The spin-fluctuation $d$-wave pairing induced by the hopping parameter $t$ was obtained within the diagram technique for the HOs in the  $t$--$J$ model~\cite{Izyumov91,Izyumov92}.

A number of studies of the  $t$--$J$ model at low doping  predict that doped holes dressed by strong AFM spin fluctuations propagate coherently as quasiparticle spin-polarons with a narrow band of the order of $J$ (see, e.g., ~\cite{Martinez91,Liu92}).  In Ref.~\cite{Plakida97} the singlet superconducting pairing of  spin-polarons on the AFM background  was found.

The memory function approach was used in Refs.~\cite{Prelovsek02,Sega03,Prelovsek04,Prelovsek06,Sega06} to study the magnetic susceptebility in cuprates within the $t$--$J$ model and to explain the emergence of the magnetic resonance mode.  A theory of spin excitations  within the relaxation-function approach  for the dynamical spin susceptibility in the $t$--$J$ model was developed  in the normal~\cite{Vladimirov09} and superconducting~\cite{Vladimirov11} states. It was shown that the magnetic resonance mode is caused  by a weak damping of the spin excitations close to the AFM wave vector and does not relate to the superconducting phase transition contrary to the theoretical description based on the spin-exciton model~\cite{Sega03,Onufrieva02,Eremin05}.

A possibility of HTSC mediated  by AFM spin fluctuations as a ``glue'' for
superconducting pairing  was considered within phenomenological spin-fermion models in many publications (see, e.g.,~\cite{Scalapino95,Monthoux94,Moriya00,Abanov03,Chubukov04,Abanov08} and
references therein). In the spin-fermion models the conventional Fermi-liquid approach was used where the kinematical interaction of electron with spin fluctuation  is lost.  In the theory a fitting parameter for electron interaction with spin excitations has to be introduced contrary to the microscopic theory~\cite{Plakida99,Prelovsek05} where this interaction is given by the hoping parameters.

In  our previous publications~\cite{Plakida07,Plakida13,Plakida14,Plakida16} we study the electronic spectrum and superconductivity in  the 2D extended Hubbard model using equation of motion method for the GFs.
In the present paper we consider superconductivity in  the limit of
strong correlations, $U \gg t$, using the
extended  $t$--$J$--$V$ model  with an  intersite Coulomb interaction (CI)
$V_{ij}$ and the electron-phonon interaction (EPI) generalizing our previous results for the conventional $t$--$J$ model~\cite{Plakida99}. Consideration of EPI and CI permits to compare the electron-phonon pairing mechanism with the spin-fluctuation one and to estimate the role of CI in suppression  of the superconducting $T_c$.

We derive the Dyson equations for the normal and anomalous GFs in  terms  of the HOs where the self-energy is calculated in the self-consistent Bohr approximation (SCBA). At first we consider the generalized MFA (GMFA) where the exchange interaction $J$ and CI $V_{ij}$ determine the electronic spectrum and the Fermi surface (FS) for the well-defined qausiparticle (QP) excitations. Taking into
account the self-energy effects the electronic spectral functions, the damping of QP excitations and the FS as functions of doping are calculated.  At low doping the arc-type FS is  emerging. Analyzing the gap equation  we show that the  strong kinematical interaction of electrons  with AFM
spin-fluctuations results in the $d$-wave superconductivity with high-$T_c$.  Contribution from the EPI  to the $d$-wave pairing turns  out to be small but it determines a weak isotope effect.

In the next Section we present the general formulation of the model and derivation of the Dyson equation. In Section~\ref{sec:3} the  GMFA for the normal and superconducting states is considered. The self-energy calculation is given in  Section~\ref{sec:4}.
The results and discussion are presented in Section~\ref{sec:5}. Summary is given in  Section~\ref{sec:6}. Details of calculations are shown in Appendix.

\section{General formulation}
\label{sec:2}

\subsection{Extended t-J-V  model }

\label{sec:2.1}

We consider electronic spectrum and superconducting pairing in the extended $t$--$J$-$V$  model on a square lattice.  To study strong electron correlations in the singly occupied subband of the $t$--$J$ model  one has to use the projected electron operators, as $\widetilde{a}_{i \sigma}^{\dag}  = a_{i \sigma}^{\dag}  (1-N_{i\bar{\sigma}})$.  Here $\, a_{i \sigma}^{\dag} $ is a creation electron operator on the lattice site $i$  with spin $\sigma/2, \; \sigma=\pm 1 \; (\bar{\sigma} = - \sigma)\,$ and  $ N_{i\bar{\sigma}} = \widetilde{a}_{i\bar{\sigma}}^{\dag} \widetilde{a}_{i\bar{\sigma}}$ is the number operator. The $t$--$J$ model in the conventional notation reads:
\begin{eqnarray}
H = -\sum_{i \neq j,\sigma}t_{ij} \widetilde{a}_{i \sigma}^{+} \widetilde{a}_{j \sigma} +
\frac{1}{2} \sum_{i \neq j} \, J_{ij} \left ( \textbf{S}_{i} \textbf{S}_{j}
- \frac{1}{4} N_{i} N_{j} \right ) + H_{c, ep}
 \label{1}
\end{eqnarray}
where   $S_{i}^{\alpha  } = (1/2) \sum_{s,s'} \widetilde{a}_{is}^{+}
\sigma_{s,s'}^{\alpha} \widetilde{a}_{is'}$  are spin-$1/2$ operators, $\sigma_{s,s'}^{\alpha}$ is the Pauli matrix.  Here $t_{ij}$ is the hopping parameter between $i$ and $j$ lattice sites  and $J_{ij}$ is the AFM  exchange interaction.  The intersite CI  $V_{ij}$ for electrons and EPI   $g_{ij}$ are taken into account   by the Hamiltonian:
\begin{eqnarray}
H_{c, ep} &= & \frac{1}{2}  \sum_{i\neq j}\,V_{ij} N_i N_j + \sum_{i, j}\,g_{i j} N_i\,
u_j,
 \label{1a}
\end{eqnarray}
where $u_j$ describe  atomic displacements on the lattice site $j$ for phonon modes.

 The unconventional commutation relations for the projected electron operators result in the kinematical interaction. For instance,  if we consider commutation relation for the projected electron creation $\widetilde{a}_{j \sigma}^{\dag}\, $ and annihilation $  \widetilde{a}_{i \sigma} \,$  operators,
\begin{eqnarray}
 \widetilde{a}_{i \sigma} \widetilde{a}_{j \sigma}^{\dag}  +
 \widetilde{a}_{j \sigma}^{\dag} \widetilde{a}_{i \sigma}
 = \delta_{ij}(1 - N_{i \sigma} /2 + \sigma S_i^z),
 \label{2}
\end{eqnarray}
we observe that they  are Fermi operators on different lattice sites but on the same lattice site they describe the kinematical interaction of electrons with charge  $N_{i\sigma}$ and spin $S_i^\alpha$ fluctuations.

It is convenient to describe the projected electron operators by the Hubbard operators
(HOs)~\cite{Hubbard65}, as, e.g., $\widetilde{a}_{i \sigma}^{+} = X_{i}^{\sigma 0}$.
Using the HOs,  we write the Hamiltonian  (\ref{1}) in the form
\begin{eqnarray}
H &=&
 - \sum_{i \neq j,\sigma}t_{ij}X_{i}^{\sigma 0}X_{j}^{0\sigma}
 - \mu \sum_{i \sigma} X_{i}^{\sigma \sigma}
 \nonumber\\
&+& \frac{1}{4} \sum_{i \neq j,\sigma} J_{ij}
\left(X_i^{\sigma\bar{\sigma}}X_j^{\bar{\sigma}\sigma}  -
   X_i^{\sigma\sigma}X_j^{\bar{\sigma}\bar{\sigma}}\right)
   + H_{c, ep} ,
   \label{3}
\end{eqnarray}
where the HOs  $X_{i}^{\alpha\beta} = |i\alpha\rangle\langle i\beta|$ describe
transitions from the state  $|i,\beta\rangle$ to the state $|i,\alpha\rangle$ on the
lattice site $i$ for  three electronic states: the unoccupied state $(\alpha, \beta =0) $ and two singly occupied states $(\alpha, \beta = \sigma)$. The chemical potential  $\mu$ in (\ref{3}) is determined  from the equation for the average number of electrons:
\begin{equation}
  n =   \langle \, N_i \rangle ,
    \label{4}
\end{equation}
where   $\langle \ldots \rangle$  is the statistical average with the Hamiltonian
(\ref{3}).

The number and  spin operators in the HO representation read
\begin{eqnarray}
N_{i}  &=& \sum_{\sigma }N_{i\sigma}  ,\quad N_{i\sigma} = X _{i}^{\sigma \sigma} ,
\label{5a}\\
S_{i}^{\sigma} & = & X_{i}^{\sigma\bar\sigma} ,\quad
 S_{i}^{z} =  (\sigma/2) \,( X_{i}^{\sigma \sigma}  -
  X_{i}^{\bar\sigma \bar\sigma}) .
\label{5b}
\end{eqnarray}
The HOs satisfy the completeness relation $\,
  X_{i}^{00} + X_{i}^{\sigma\sigma} +
  X_{i}^{\bar\sigma\bar\sigma}  = 1 \,$,
which shows that only one quantum state $|i,\alpha\rangle $ on each lattice site $i$ can be occupied and, therefore, rigorously preserves the constraint of no double occupancy.  From the multiplication rules for the HOs $\, X_{i}^{\alpha\beta}X_{i}^{\gamma\delta}=\delta_{\beta\gamma} X_{i}^{\alpha\delta}$  for Fermi-type operators $X_{i}^{0\sigma}$ follow  the commutation relations as in Eq.~(\ref{2})
for  $  \widetilde{a}_{i \sigma}$, while for  Bose-type operators such as the number  (\ref{5a}) or the spin (\ref{5b}) operators the commutation relations read:
\begin{equation}
\left[X_{i}^{\alpha\beta} , X_{j}^{\gamma\delta}\right] =X_{i}^{\alpha\beta} X_{j}^{\gamma\delta} - X_{j}^{\gamma\delta} X_{i}^{\alpha\beta}=
\delta_{ij}\left(\delta_{\beta\gamma}X_{i}^{\alpha\delta} -
\delta_{\delta\alpha}X_{i}^{\gamma\beta}\right).
 \label{6}
\end{equation}
These  commutation relations determine the kinematical interaction for the HOs.

\subsection{ Dyson equation}

\label{sec:2.2}

To discuss the electronic spectrum and superconducting pairing within the model (\ref{1}) we consider the  matrix GF~\cite{Zubarev60}
\begin{eqnarray}
\widehat{G}_{i j,\sigma} (t-t^{\prime}) & = &
-i \theta(t-t')\langle \, \{ \, \Psi_{i \sigma}(t) ,
  \Psi_{j \sigma}^+ (t^{\prime}) \, \} \, \rangle
 \nonumber \\
 & \equiv &  \langle  \langle \Psi_{i \sigma}(t)\, |
 \, \Psi_{j \sigma}^+ (t^{\prime}) \rangle\rangle
\label{7},
 \end{eqnarray}
where $\theta(x) $ is the Heviside function$, \{A, B\} = AB + BA$,  $ A(t)= \exp (i Ht) A\exp (-i Ht)$ $(\hbar = 1)$, and we introduced  HOs in the  Nambu notation:
\begin{equation}
\Psi_{i \sigma} =  { X_i^{0 \sigma } \choose
 X_i^{ \bar{\sigma} 0} }
,\qquad \Psi_{i \sigma}^+ = \left( X_i^{ \sigma 0} \; X_i^{0 \bar{\sigma}} \right)\; .
\label{7a}
\end{equation}
The Fourier representation in ($\bf {k},\omega $)-space is defined by the relations:
\begin{eqnarray}
\widehat{G}_{ij\sigma } (t-t') =\frac{1}{2\pi } \int_{-\infty }^{\infty } dt e^{-i\omega (t-t')} \widehat{G}_{ij\sigma}(\omega),
\label{8a} \\
\widehat{G}_{ij\sigma} (\omega ) = \frac{1}{N} \sum_{\bf k} \exp [{\bf k} ({\bf r_i- r_j}) ] \,\widehat{G}_{\sigma} (\bf {k},
\omega) ,
 \label{8b}
\end{eqnarray}
where $N$ is the number of lattice sites.
The  GF (\ref{7})  is convenient to write in the matrix form
\begin{equation}
\widehat{G}_{\sigma}({\bf k}, \omega)=
  { G_{\sigma}({\bf k}, \omega)  \quad \quad
  F_{\sigma}({\bf k}, \omega) \choose
 F_{\sigma}^{\dagger}({\bf k}, \omega) \quad
   -{G}_{\bar\sigma}(-{\bf k}, -\omega)} ,
 \label{7c}
\end{equation}
where $G_{\sigma}({\bf k}, \omega)$  and $
  F_{\sigma}({\bf k}, \omega)$   are the normal and anomalous parts of the GF (\ref{7}) .

To calculate  GF (\ref{7}) we use the projection technique in the equation of motion method~\cite{Plakida12}. By differentiating the
GF over the time $t$ we get  the following equation
\begin{equation}
\omega \widehat{G}_{ij\sigma} (\omega) = \delta_{ij} \widehat{Q}_{\sigma} +
\langle\!\langle \widehat{Z}_{i\sigma} \mid  \Psi_{j \sigma}^+ \rangle\!\rangle_{\omega},
 \label{9}
\end{equation}
where $\widehat{Z}_{i\sigma} = [\Psi_{i\sigma}, H]  $.
The matrix $\widehat{Q}_{\sigma}$ is the average value of  time-independent operators $ \widehat{Q}_{\sigma}=  \langle \{ \Psi_{i\sigma}, \Psi_{j \sigma}^{+} \}\rangle $. The  diagonal  matrix element is given by
$\, \langle \{ X_i^{0 \sigma } , X_i^{ \sigma 0}\}\rangle = \langle  X_i^{ 0 0} + X_i^{ \sigma \sigma } \rangle  = 1 - ({1}/{2}) \langle N_{i \sigma} \rangle + \langle S_i^z \rangle $, while the off-diagonal matrix element  $\,\langle  \{ X_i^{0 \sigma } , X_i^{0 \bar{\sigma}}\} \rangle = 0$.  Thus, in the paramagnetic state, $\langle S_i^z \rangle  = 0$,  the matrix
\begin{equation}
 \widehat{Q}_{\sigma}  = \hat \tau_{0}  Q,\;  Q =   1-n/2 ,
\label{9a}
\end{equation}
is the  $\sigma$-independent unity matrix determined by the average number of electrons (\ref{4}). Therefore, in the following equations this matrix can be replaced by the scalar $Q$.

Now, we project the many--particle GF in (\ref{9}) on the single--electron GF
\begin{eqnarray}
\langle\!\langle \widehat{Z}_{i\sigma} \mid  \Psi^+_{j\sigma} \rangle\!\rangle_{\omega}
= \sum_l \widehat{E}_{il\sigma} \langle\!\langle \Psi_{l\sigma}\mid
\Psi^+_{j\sigma}\rangle\!\rangle_{\omega}
+ \langle\!\langle \widehat{Z}_{i\sigma}^{(irr)} \mid
\Psi^+_{j\sigma} \rangle\!\rangle_{\omega},
\label{10}
\end{eqnarray}
where we introduce  the {\it irreducible}  part of the operator $\widehat{Z}_{i\sigma}$:
\begin{eqnarray}
\langle \, \{ \widehat{Z}^{(irr)}_{i\sigma}, \Psi^{+}_{j\sigma} \} \,\rangle = \langle
\widehat{Z}^{(irr)}_{i\sigma} \Psi^{+}_{j\sigma} + \Psi^{+}_{j\sigma}
\widehat{Z}^{(irr)}_{i\sigma}   \rangle =0 .
\label{10a}
\end{eqnarray}
This  results in the equation for the matrix of electronic energy in the GMFA:
\begin{equation}
\widehat{E}_{ij\sigma} = \langle \{\, [\Psi_{i\sigma}, H], \Psi^+_{j\sigma}\} \rangle \;
{Q}^{-1}   =\left( \begin{array}{cc}
 \varepsilon_{ij}  & \Delta_{ij\sigma }     \\
 \Delta_{ji\sigma}^{*} & -\varepsilon_{ji}
\end{array}\right),
\label{11}
\end{equation}
which in  the Fourier representation reads:
\begin{equation}
{\widehat E}_{\sigma}({\bf k}) = \frac{1}{N} \sum_{\bf r_i, r_j} {\rm e}^{-
 {\bf k} ({\bf r_i- r_j})} \, \widehat{E}_{ij\sigma } = \left(
\begin{array}{cc}
 {\varepsilon}({\bf k})   & {\Delta}_{\sigma}({\bf k})  \\
   {\Delta}^{*}_{\sigma}({\bf k}) &
     -{\varepsilon}({\bf k})
\end{array}\right).
 \label{11a}
\end{equation}
Here $\, {\varepsilon}({\bf k}) $  is  the electronic spectrum in the normal state and ${\Delta}_{\sigma}({\bf k}) $ is the gap in the superconducting state.
The energy matrix (\ref{11a}) defines the zero--order GF:
\begin{equation}
{\widehat G}^0_{\sigma}({\bf k},\omega) = Q  \frac{\omega\hat \tau_{0}+
{\varepsilon}({\bf k}) \hat \tau_{3}+ \Delta_{\sigma}({\bf k}) \hat \tau_{1} }
{\omega^{2} - E^2({\bf k}) },
\label{12}
\end{equation}
where $ \hat \tau_{1}, \; \hat \tau_{3} $ are the  Pauli matrices and $E^2({\bf k}) = {\varepsilon}^{2}({\bf k}) +
\Delta^{2}_{\sigma}({\bf k})$ is  the energy of quasiparticle (QP) excitations in the superconducting state.

By writing the equation of motion
for the   irreducible part of the GF in~(\ref{10}) $\, \langle\!\langle \widehat{Z}_{i\sigma}^{(irr)} (t) \mid \Psi^+_{j\sigma} (t') \rangle\!\rangle$  with respect to the second time
$t^{\prime}$ for the right--hand side operator $\Psi^+_{j\sigma} (t^{\prime})$ and
performing the same projection procedure as in (\ref{10}) we can obtain the Dyson
equation for the GF (\ref{7}) in the form
\begin{equation}
\widehat{G}_{ij\sigma} (\omega) = \widehat{G}^0_{ij\sigma}(\omega) + \sum_{kl}
\widehat{G}^0_{ik\sigma} (\omega)\ Q^{-1}\,\widehat{\Sigma}_{kl\sigma}(\omega) \
\widehat{G}_{lj\sigma} (\omega) .
\label{15}
\end{equation}
The self--energy operator $\widehat{\Sigma}_{kl\sigma} (\omega)$   is given by the {\it proper}
part of the scattering matrix that  has no parts connected by the single zero-order GF (\ref{12}):
\begin{equation}
\widehat{\Sigma}_{ij\sigma} (\omega) = \langle\!\langle
\widehat{Z}^{(irr)}_{i\sigma} \mid \widehat{Z}^{(irr)^+}_{j\sigma}
\rangle\!\rangle_{\omega}^{\rm proper}  {Q}^{-1}   \; .
 \label{16}
\end{equation}

The self-energy operator can be  written in the same matrix form as the  GF (\ref{7c}):
\begin{equation}
 \widehat{\Sigma}_{\sigma}({\bf k}, \omega) =  {M_{\sigma}({\bf k}, \omega) \quad \quad
\Phi_{\sigma}({\bf k}, \omega) \choose \Phi_{\sigma}^{\dagger} ({\bf k}, \omega)\quad
- {M}_{\bar\sigma}({\bf k}, -\omega)} \, ,
 \label{16a}
\end{equation}
where the  $M_{\sigma}({\bf k}, \omega)$ and $\Phi_{\sigma}({\bf k}, \omega) $  denote the respective normal and anomalous (pair) components of the self-energy operator.
Therefore, for   the single--electron GF (\ref{7}) we obtain an exact representation:
\begin{equation}
\widehat{G}_{ \sigma} ({\bf k},\omega) = {Q} \{\omega \hat \tau_0  - \widehat{E}_{\sigma}({\bf k}) -  \widehat{\Sigma}_{\sigma}({\bf k}, \omega) \}^{-1} .
 \label{17}
\end{equation}
A formal solution of the matrix equation (\ref{17}) can be written in the form  (cf.~\cite{Eliashberg60}):
\begin{eqnarray}
&& {\widehat G}_{\sigma}({\bf k},\omega)
 \nonumber\\
&& = Q  \frac{\omega Z_{{\bf k}}(\omega)\hat
\tau_{0}+({\varepsilon}({\bf k}) +\xi_{{\bf k}}(\omega))\hat\tau_{3}
+ \phi_{\sigma}({\bf k}, \omega)\hat \tau_{1} }
 {(\omega Z_{{\bf k}}(\omega))^{2}-({\varepsilon}({\bf k})+
 \xi_{{\bf k}}(\omega))^{2} - |\phi_{\sigma}({\bf k},\omega)|^{2} },
\label{17a}
\end{eqnarray}
where we introduced the odd and even components of the
normal self-energy operator $\,  M_{\sigma}({\bf k}, \omega) $ with respect to the frequency $\omega$:
\begin{eqnarray}
\omega(1-Z_{\bf k}(\omega))&=&\frac{1}{2}[ M({\bf k},\omega) -  M(-{\bf k},-\omega)]  ,
\label{17b} \\
\xi_{\bf k}(\omega)& = &\frac{1}{2}[ M({\bf k},\omega) +  M(-{\bf k},-\omega)] .
\label{17c}
\end{eqnarray}
The superconduction gap $\phi_{\sigma}({\bf k}, \omega)$   is determined both by the GMFA function $\Delta_{\sigma}({\bf k}) $ in Eq.~(\ref{11a})  and the anomalous self-energy component $\Phi_{\sigma}({\bf k},\omega)$ in Eq.~(\ref{16a}) :
 \begin{eqnarray}
 \phi_{\sigma}({\bf k}, \omega)& = & \Delta_{\sigma}({\bf k})  +\Phi_{\sigma}({\bf k},\omega) .
\label{17d}
\end{eqnarray}

The QP excitation in the GMFA  (\ref{11a}) is determined by the static correlation functions and can be directly calculated as described in the next section.  However, to calculate the self-energy matrix (\ref{16a}) which describes inelastic scattering of electrons on spin,  charge fluctuations and phonons one has to introduce an approximation for the many--particle GFs in (\ref{16a}) as considered in Section~4.

\section{ Generalized mean-field approximation }

\label{sec:3}

\subsection{ Normal state }

\label{sec:3.1}

The normal state GF in the GMFA is given by Eq.~(\ref{12})  for the zero gap function:
 \begin{equation}
 G^0({\bf k},\omega) = \langle\langle X_{\bf k}^{0 \sigma} |
  X_{\bf k}^{\sigma 0 } \rangle  \rangle_{\omega} = \frac{Q}{\omega - \varepsilon({\bf k}) } .
\label{12a}
\end{equation}
To calculate the energy $\varepsilon({\bf k})$  we use the equation of motion for the HOs
\begin{eqnarray}
\left( i\frac{d}{dt} +\mu \right) X_{i}^{0\sigma} =
 -  \sum_{j, \sigma'} t_{ij} B_{i \sigma \sigma'} X_{j}^{0\sigma'}
+ X_{i}^{0\sigma} \sum_{ j}\,g_{i j} u_j
\nonumber\\
 + \frac {1}{2} \sum_{j, \sigma'} J_{ij} (B_{j \sigma \sigma'}-\delta_{\sigma \sigma'})
X_{i}^{0\sigma'}+ X_{i}^{0\sigma}\sum_{j} V(i,j)\,N_{j} .
\label{13}
\end{eqnarray}
 Here  we introduced the Bose-like operator
\begin{eqnarray}
 B_{i\sigma\sigma'} & = &
     (X^{00}_{i} + X^{\sigma\sigma}_{i})\delta_{\sigma'\sigma}
  +   X^{\bar{\sigma}\sigma}_{i}\delta_{\sigma' \bar{\sigma}}
\nonumber\\
&=& (1- \frac{1}{2} N_i + \sigma S^z_i) \delta_{\sigma'{\sigma}}
 + S^{\bar{\sigma}}_{i} \delta_{\sigma' \bar{\sigma}} ,
\label{14}
\end{eqnarray}
which describes electron scattering on spin and charge fluctuations caused by the kinematic interaction~(\ref{2}).
Using Eq.~(\ref{13}) we calculate   the matrix  $\varepsilon_{ij} = \langle \{ [X_i^{0\sigma}, H],X_j^{\sigma 0}  \} \rangle \;
{Q}^{-1}$ and for the electronic energy  ${\varepsilon}({\bf k})$ obtain the relation:
\begin{eqnarray}
 {\varepsilon}({\bf k}) & = & -4 t\,\alpha \gamma({\bf k})
 - 4 t'\,\beta \gamma'({\bf k})
- 4 t'' \,\beta \gamma''  ({\bf k})
\nonumber\\
 & - &\frac{2J}{N}\sum_{q} \gamma \left ( k-q \right ) N_{q\sigma } + \omega^{(c)}({\bf k})
  -  \mu  ,
\label{18}\\
 \omega^{(c)}({\bf k}) &  = & \frac{1}{N } \sum_{\bf q}
  V({\bf k-q}) N({\bf q}) ,
\label{18a}
\end{eqnarray}
where the renormalization of the chemical potential $\delta \mu $ in the GMFA we include in the definition of $\mu$. Here  $t, \, t' , \, t''$ are the hopping parameters  between the first  $\,{\bf a}_1 =\pm a_x, \pm a_y \,$, second
$\, {\bf a}_2 =\pm (a_x \pm a_y)$, and   third $\, {\bf a}_3 =\pm 2a_x, \pm 2a_y\, $ neighbors, respectively ($ a_x = a_y= a\,$ - are the  lattice constants).
The Fourier components of the hopping
parameter $ t({\bf k })$,   CI $ V({\bf k})$ and the exchange interaction $J({\bf k}) $
are given by:
\begin{eqnarray}
 t({\bf k }) & = &  4\, t \, \gamma ({\bf k }) + 4\, t' \, \gamma' ({\bf k })
+ 4 t'' \, \gamma'' ({\bf k }),
\label{4a} \\
 V({\bf k }) & = & 4 V_1 \, \gamma ({\bf k }) + 4 V_2 \, \gamma'({\bf k}) ,
\label{4b}  \\
  J({\bf k }) & = & 4 J \gamma ({\bf k }),
\label{4c}
\end{eqnarray}
where $ \gamma ({\bf k }) = ({1}/{2}) (\cos k_x + \cos k_y ),\;
 \gamma'({\bf k})= \cos k_x  \cos k_y , \;
 \gamma ''({\bf k }) = ({1}/{2}) (\cos 2k_x + \cos2 k_y )$.  For the intersite CI for the
first and the  second neighbors, $V_1$ and $V_2 $ in (\ref{4b}),  we take sufficiently small values $\, V_1 = 0.3\, t\;$  and  $\, V_2 =   0.2 \, t  \,$ as shown in  numerical calculations~\cite{Feiner96}. For the AFM exchange interaction (\ref{4c}) we take   $\, J = 0.4 t\,$. Below we take $t = 0.4$~eV as the energy unit and put $t = 1$.

The renormalization of   the spectrum (\ref{18}) caused by the AFM short-range correlations is determined  by the parameters:
\begin{eqnarray}
 \alpha &= & Q \, \Big( 1 + {C_{1}}/{Q^2}\Big), \;
 \beta = Q\, \Big( 1 + {C_{2}}/{Q^2}\Big),
 \label{19}
\end{eqnarray}
where the spin correlation functions for the first and the next neighbors are:
\begin{eqnarray}
C_{1} &=& \langle {\bf S}_i{\bf S}_{i\pm a_{x}/a_{y}} \rangle = \frac{1}{N} \sum_{\bf q}
\gamma({\bf q})\,  C_{\bf q},
\nonumber \\
 C_{2} & = & \langle {\bf S}_i{\bf S}_{i\pm a_{x}\pm a_{y}} \rangle =
\frac{1}{N} \sum_{q} \gamma'({\bf q})\,  C_{\bf q} .
 \label{19a}
\end{eqnarray}
For  the  spin  correlation function $\, C_{\bf q}=\langle {\bf S}_{\bf q}{\bf S}_{-\bf
q} \rangle $ we take the model:
\begin{eqnarray}
C_{\bf q} = \frac{C_{\bf Q}}{1+\xi^2[1+ \gamma({\bf q})]},
 \label{19b}
\end{eqnarray}
where the parameter $ C_{\bf Q}  $ is defined from the normalization condition $\langle
{\bf S_i S_i }\rangle =  (3/4) n = (1/N)\sum_{\bf q} C_{\bf q} $. The correlation functions have the maximum $ C_{\bf Q}  $ at the AFM wave vector ${\bf Q} = (\pi, \pi)$. In Ref.~\cite{Vladimirov09}  the correlation functions $C_1$ and $C_2$ as a function of doping were calculated for the $t-J$ model  (see Fig.~1). To simplify the numerical calculations  it is more convenient to use the  analytical equations (\ref{19a}),  (\ref{19b}) which give  $C_1$ and $C_2$  close to that ones in Ref.~\cite{Vladimirov09}. The correlation functions  depend on the AFM correlation length $\xi$.
 For its dependence on the hole doping  $ \delta$  we use an approximation $\xi/a  = 1/\sqrt{\delta}$  observed in neutron scattering experiments (see, e.g.,~\cite{Birgenau88})
and confirmed in the exact diagonalization study for finite clusters \cite{Bonca89}.
The values of correlation functions $C_1, \, C_2,\, C_{\bf Q}  $, the AFM correlation length  $\xi$, and
parameters $\alpha, \,\beta$ for various hole doping  $ \delta$ are given in
Table~\ref{Table1}. The value of the normalization parameter ${\chi}_{{\bf Q}}= { 2C}_{{\bf Q}}/\omega_s $, $\,\omega_s  = J = 0.4$,  for the dynamical spin susceptibility in Eq.~(\ref{r1a}) is also given.

\begin{table}
\centering \caption{Static spin correlation  functions $C_1, \, C_2 \,$ and
renormalization parameters $ C_{\bf Q}$ and  ${\chi}_{{\bf Q}}  $  for the AFM correlation length  $\xi/a  = 1/\sqrt{\delta}$  at various hole concentrations
$ \delta = 1 - n$.}
\label{Table1}      
 \begin{tabular}{lrrrrrr}
\noalign {\bigskip}
 \noalign{\smallskip}\hline
\noalign{\smallskip}
 \quad $\delta =$  \quad    & 0.05  & 0.10  &  0.20 & 0.30  & 0.40 \\
 \noalign{\smallskip} \hline\\
  \quad $ {\xi}= 1/\sqrt{\delta}$    & 4.5 & 3.2  & 2.2 & 1.8 & 1.6 \\
    \quad ${C}_1({\xi})$    &-0.3 &-0.24  &-0.17 &-0.13  & -0.1 \\
 \quad ${C}_2({\xi}) $  & 0.2 & 0.14  & 0.09& 0.06  & 0.04 \\
 \quad ${C}_{{\bf Q}}({\xi}) $   &  9.00 & 5.02 &  2.74 & 1.84 & 1.33 \\
 \quad ${\chi}_{{\bf Q}}({\xi}) $   &44.98 & 25.11 & 13.68 & 9.21 &6.64 \\
  \quad $\alpha({\xi})$    & -0.043 &0.11   & 0.31 &0.45  & 0.56 \\
 \quad $\beta({\xi}) $  & 0.9 & 0.8  & 0.75 & 0.74  & 0.76 \\
  \noalign{\smallskip}
 \hline\noalign{\smallskip}
\end{tabular}
\end{table}

The electronic occupation number in the GMFA is determined by the zero-order GF (\ref{12a})
\begin{equation}
 N({\bf k}) =
 \frac{Q}{ \exp[ {\varepsilon}({\bf k}) / T] +1 }.
\label{20}
\end{equation}
The chemical potential  in the GMFA is calculated from the equation:
\begin{equation}
 n  = \frac{1}{N}\sum_{{\bf k}, \sigma} N({\bf k}) = \frac{2- n}{N}\sum_{{\bf k}}
 \frac{1}{ \exp[ {\varepsilon}({\bf k}) / T] +1 }.
\label{21}
\end{equation}
 Eq.~(\ref{21})  proves  that  $n \leq 1 $ in the singly-occupied band in the $t$--$J$ model.

\begin{figure}[htp]
 \centering
 \includegraphics[scale=0.6]{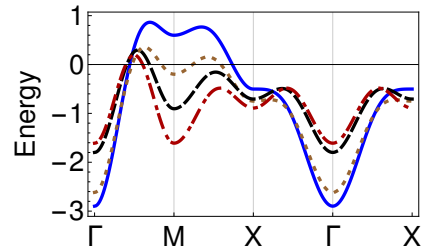}
 \caption{(Color online) Dispersion of the electron spectrum along the main
directions in the BZ: $\Gamma (0, 0) \rightarrow M (\pi,\pi) \rightarrow  X (\pi, 0)
\rightarrow \Gamma  (0, 0) \rightarrow  X (\pi, 0) \,$ for $\delta = 0.05$  (red, dash-dotted line),  $\delta = 0.1$  (black, dash line),  $\delta = 0.2$ (brown, dotted line), $\delta = 0.3$
 (blue, solid line). }
\label{fig1}
\end{figure}

\begin{figure}[htp]
\centering
\includegraphics[scale=0.5]{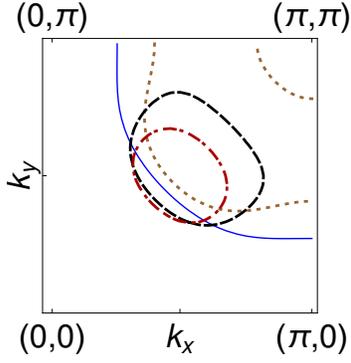}
\caption{(Color online)  Fermi surface in the quarter of the BZ for $\delta = 0.05$  (red, dash-dotted line),  $\delta = 0.1$  (black, dash line),  $\delta = 0.2$ (brown, dotted line), $\delta = 0.3$
 (blue, solid line)}.
  \label{fig2}
 \end{figure}

\begin{figure}[htp]
\centering
\includegraphics[scale=0.6]{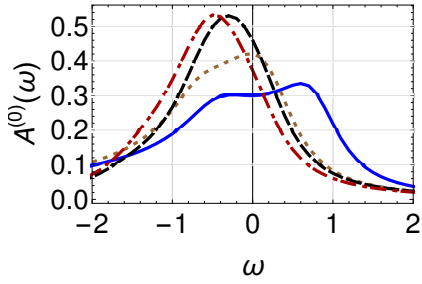}
\caption{(Color online)  Density of states $A^0(\omega)$ for $\delta = 0.05$  (red, dash-dotted line), $ \delta = 0.1$  (black, dash line), $ \delta = 0.2$ (brown, dotted line), $\delta = 0.3 $   (blue, solid line).}
\label{fig3}
\end{figure}

To reproduce the realistic electronic spectrum  which shows the FS  transition from the four-pockets at small doping  to a large one with doping for the hopping parameter   we take  $ t'= 0.1 t, \, t''=  0.2 t$. For these parameters we obtain the electronic spectrum  in
Fig.\ref{fig1} which is similar to  calculated within the Hubbard model in
Refs.~\cite{Plakida07,Plakida14}.  Note that at small doping $\delta = 0.05$ the electronic energy ${\varepsilon}({\bf k})$  at the  $\Gamma (0,0)$  and
 $ M(\pi,\pi)$ points of the BZ are close induced by short-range AFM correlations as in the long-range AFM state.
This renormalization of the spectrum results in the FS ${\varepsilon}({\bf k_{\rm F}}) =0\, $  with four hole pockets in Fig.\ref{fig2} at low doping.  As discussed  in Section~\ref{sec:4.1}, by taking into account the self-energy contribution in the GF~(\ref{17a}) instead of the well defined in the GMFA electronic spectrum   in Fig.\ref{fig1} we observe a diffuse spectral density.  At the same time,  the FS in Fig.\ref{fig2} in the form of closed pockets for low doping transfers to open arcs where
 only the outer part of pockets is revealed while the inner part,  closer to the $(\pi,\pi)$ point of the BZ,  is smoothed away.
 The  density of states (DOS) $A^0(\omega)$ in the GMFA  is determined by the relation:
\begin{equation}
A^0(\omega) =  \frac{1}{N } \sum_{\bf k} \frac{1}{\pi Q} [- {\rm Im} G^0({\bf k},\omega)]=
\frac{1}{N } \sum_{\bf k} \delta[{\varepsilon}({\bf k})- \omega],
\label{22}
\end{equation}
and presented in Fig.\ref{fig3}    as a function of doping in units of $1/t$.  With doping AFM correlations are suppressed which results in increasing of the effective bandwidth, while  the  density of state at the FS is decreasing.

\subsection{ Superconducting state }
\label{sec:3.2}

The superconducting gap function  $\Delta_{\sigma}({\bf k})$ in the zero-order GF (\ref{12}) is determined by the Fourier component of the  anomalous correlation function in the energy matrix (\ref{11}): $\Delta_{ij\sigma}
 = \langle \{ [X_i^{0\sigma}, H],X_j^{0 \bar{\sigma} }  \} \rangle \;{Q}^{-1}$.  Using the equation of motion (\ref{13}) for the gap function we find the relation
\begin{equation}
\Delta_{\sigma}({\bf k}) =  \frac{1}{N Q} \sum_{{\bf q}} \;[2 t({\bf q}) -  J({\bf k- q})  + V({\bf k- q})]
 \langle X_{- \bf q}^{0 \bar{\sigma}}  X_{\bf q}^{0 \sigma} \rangle.
\label{23}
\end{equation}
The EPI gives no contribution in the GMFA. From the  GF $F^{(0)}_{\sigma}({\bf q},\omega)$ in Eq.~(\ref{12}) we obtain for  the  correlation function
\begin{eqnarray}
\langle X_{- \bf q}^{0 \bar{\sigma}}  X_{\bf q}^{0 \sigma} \rangle  =
- Q\, \frac{\Delta_{\sigma}({\bf q})}{2  E({\bf q})}
 \tanh\frac{ E({\bf q})}{2T}.
  \label{25}
\end{eqnarray}
There are two pairing contributions in Eq.~(\ref{23}): the exchange interaction $J({\bf k -q})$  and the ${\bf k}$-independent kinematic interaction $t({\bf q})$. The latter gives ${\bf k}$-independent gap
\begin{equation}
\Delta_{\sigma} = - \frac{2}{N } \sum_{{\bf q}} \; t({\bf q})
\frac{\Delta_{\sigma}({\bf q})}{2  E({\bf q})}
 \tanh\frac{ E({\bf q})}{2T},
\label{23a}
\end{equation}
which  violates the constraint of no double occupancy (see Refs.~\cite{Plakida89,Plakida14}). In particular,
 \begin{equation}
 \langle X_i^{0\sigma}\, X_i^{0 \bar{\sigma} } \rangle=  \frac{Q}{N } \sum_{{\bf q}} \;
\frac{\Delta_{\sigma}}{2  E({\bf q})}
 \tanh\frac{ E({\bf q})}{2T} \neq 0.
\label{23a}
\end{equation}
The constraint  $\langle X_i^{0\sigma}\, X_i^{0 \bar{\sigma} } \rangle = 0$  is fulfilled for the $d$-wave gap $\Delta_{\sigma}(q_x,q_y)= - \Delta_{\sigma}(q_y,q_x)$  in integration over $(q_x,q_y)$ in (\ref{23a}) for $ E(q_x,q_y) = E(q_y,q_x)$.

\begin{figure}[htp]
\centering
\includegraphics[scale=0.5]{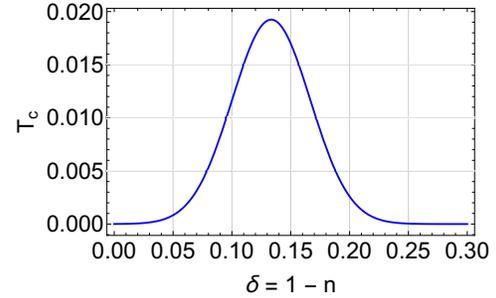}
 \caption{$T_c$ dependence on $ \delta = 1-n$ for  $V_1 = 0$. }
 \label{fig4}
 \end{figure}

Therefore, we disregard $t({\bf q})$ contribution in  Eq.~(\ref{23}) and obtain  the gap equation
 in the form
\begin{eqnarray}
\Delta_{\sigma}({\bf k}) = \frac{1}{N } \sum_{{\bf q}} \;[ J({\bf k- q})
 - V({\bf k- q})] \frac{\Delta_{\sigma}({\bf q})}{2  E({\bf q})}
 \tanh\frac{ E({\bf q})}{2T}.
\label{26}
\end{eqnarray}
Solution of the linearized  gap equation (\ref{26}) with $ E({\bf q}) = {\varepsilon}({\bf q})$ gives  $T_c$ as a function of doping for the $d$-wave pairing shown in  Fig.~\ref{fig4}. The maximal value of $T_{c}^{max}  \simeq  0.02 t \simeq 100$~K for zero CI is at  hole concentration $ \delta \sim 0.14$ for the maximal  DOS at the FS.  For conventional values of the hopping parameters without hole pockets on the FS, as, e.g., in Ref.~\cite{Plakida99},   $\,T_{c}^{max}$ is at $ \delta \sim 0.33$, far away from the experimental values  $ \delta \sim 0.16$.
Intersite CI $V({\bf k- q})$ (\ref{4b}) strongly suppresses $T_{c}$. The effective coupling for the CI $V_1 = 0.3$ results in weak coupling, $J-V_1 = 0.1$, and  extremely small  $T_{c}^{max} \sim
4 t\times 10^{-4}\,$ in comparison with with $T_{c}^{max}$  for $V_1 = 0$  in Fig.~\ref{fig4}.
Such a strong reduction of $T_c$ is explained by the unretarded character of both interactions, the exchange interaction $J$ and CI.  They   act in the whole subband of the $t-J$ model  and there is no renormalization of  CI  as in the case of retarded electron-phonon interaction. In the MFA, commonly used for the $t$--$J$ model, it is possible to explain quite high-$T_{c}  \simeq 100$~K observed in experiments as was proposed by Anderson~\cite{Anderson87}.   But taking into account the CI comparable with the exchange interaction $J$, high-$T_c$ in cuprates cannot be realized.

The three-site interaction $H_3$ in the  $t$--$J^{*}$ model further results in decreasing of the effective AFM interaction, in the simplest MFA~\cite{Yushankhai90},  $J^{*} \simeq J(n/2)$. A more accurate estimation of the three-site interaction $H_3$ in the  $t$--$J^{*}$ model in Ref.~\cite{Valkov02} has shown that the superconductivity temperature $T_{c}^{max}$ for the $d$-wave pairing  is strongly suppressed, approximately in 25 times, in comparison with the original $t$--$J$ model. So, even without CI $V_1$ the superconducting  $T_{c}^{max}$ is very small and many calculations performed in MFA for the original $t$--$J$ model, as  e.g., in Ref.~\cite{Spalek78} where a good agreement with experiments in cuprates was claimed, cannot explain  high-$T_{c}$  in cuprates.

 To find the superconductivity in this case we should take into account the kinematical interaction in the anomalous self-energy $\Phi_{\sigma}({\bf k}, \omega) $ in Eq.~(\ref{16a}) as discussed in the next Section.

\section{Self-energy calculation}
\label{sec:4}

In this section we consider the strong coupling approximation (SCA) by taking into account the self-energy contribution to the GFs in comparison to the weak-coupling approximation (WCA) in the GMFA discussed in the previous section.

The self-energy (\ref{16})  is determined by  the many-particle GFs where the  normal and anomalous (pair) components in the matrix (\ref{16a}) are given by:
\begin{eqnarray}
 M_{ij \sigma}(t - t') &=& (1/Q)\, \langle\langle [X_{i}^{0 \sigma} , H ](t)| [ H, X_{j}^{\sigma0}] (t') \rangle\rangle,
\label{29} \\
\Phi_{ij \sigma}(t - t')  &=& (1/Q)\,   \langle\langle [X_{i}^{0 \sigma} , H ](t) |
 [ X_{j}^{0\bar{\sigma}}, H](t') \rangle\rangle.
 \label{30}
\end{eqnarray}
As follows from the equation of motion (\ref{13}), the many-particle GFs in Eqs.~(\ref{29}), (\ref{30}) describe propagation of a pair of excitations, a Fermi one like $  \widetilde{a}_{i \sigma} = X_{i}^{0 \sigma} $ and a Bose one like  spin and charge fluctuations or phonons,  from time $t$ to $t'$. Their  interaction is determined by the zero order vertexes, $t_{ij}, J_{ij}$ or $ V_{ij} $. This representation of the self-energy  distinguishes from that one used in the diagram technique where the self-energy is usually given by a skeleton  diagram for a  fermion and a boson with a full vertex. This distinction results in different type of approximations for the self-energy. In the diagram technique the vertex is calculated by a perturbation theory, while for the self-energy ~(\ref{29}), (\ref{30}) we have to use an approximation for the many-particle GFs. Here we use the SCBA which is similar to the  non-crossing approximation in the diagram technique. In the approximation, a propagation for  Fermi-like  and Bose-like excitations is assumed to be independent between time $t$ to $t'$. The time-dependent many-particle correlation functions in the proper part of the self-energy, having no parts connected by a single-electron GF, therefore can be written as a product of fermionic and bosonic time-dependent correlation functions.

In particular,  the contribution from the kinematical interaction, the first sum in the
equation of motion~(\ref{13}),  is given by the decoupling of two-time correlation functions:
\begin{eqnarray}
&& \langle X_{m}^{\sigma'0}B^{+}_{j \sigma\sigma'}| X_{l}^{0\sigma'}(t)
 B_{i \sigma\sigma'}(t)\rangle
\nonumber \\
&&
 = \langle X_{m}^{\sigma' 0} X_{l}^{0\sigma'}(t)\rangle  \langle B^{+}_{j \sigma\sigma'}
   B_{i \sigma\sigma'}(t)\rangle ,
\label{B5} \\
&& \langle  X_{m}^{\bar{\sigma}' 0} B\sb{j \bar{\sigma}\bar{\sigma}'}
 | X_{l}^{\sigma' 0}(t) B\sb{i\sigma \sigma'}(t)\rangle
\nonumber \\
&& =  \langle  X_{m}^{\bar{\sigma}' 0}  X_{l}^{\sigma' 0}(t)\rangle\,
  \langle   B\sb{j\bar{\sigma}\bar{\sigma}'}
  B\sb{i\sigma \sigma'}(t) \rangle .
 \label{B6}
\end{eqnarray}
The decoupling is performed  on different lattice sites  ($ j \neq m, \; i \neq l$) which exclude   correlations between different type of excitations. The same type of approximation for the fermion-phonon GF $\langle X_{m}^{\sigma 0 } u_j | X_{l}^{0\sigma'}(t) u_i (t) \rangle =
\langle X_{m}^{\sigma 0 }  X_{l}^{0\sigma'}(t) \rangle  \langle u_j  u_i (t) \rangle  $  directly reproduces  the Eliashberg  theory~\cite{Eliashberg60} of superconductivity for  the electron-phonon model. The vertex correction to the EPI in the Eliashberg theory is  small given by the parameter $\omega_{ph}/\mu$ where $\omega_{ph}$ is a phonon energy. In the SCBA the vertex correction to the spin-fluctuation interaction is also  small, of the order of  $\omega_s/\mu$, the spin-fluctuation energy $\omega_s$ which restricts the region of interaction, to the Fermi energy $\mu$.
The time-dependent single-particle correlation functions at the  right-hand side in Eqs.~(\ref{B5}) and (\ref{B6}) are calculated self-consistently using the corresponding full GFs.

Using the spectral representation for these GFs, we obtain  the   expressions for the normal and anomalous components of the self-energy (see Appendix):
\begin{eqnarray}
  M({\bf k},\omega)  =    \frac{1}{N} \sum\sb{\bf q}
   \int\limits\sb{-\infty}\sp{+\infty}\!\! \frac{d z}{\pi Q}\,
   K^{(+)}(\omega,z,{\bf k, q}) [- \rm{Im}] \, G({\bf q},z) ,
   \label{29a}\\
\Phi\sb{\sigma}({\bf k},\omega) =
 \frac{1}{N} \sum\sb{\bf q}
   \int\limits\sb{-\infty}\sp{+\infty}\!\! \frac{d z}{\pi Q}\,
   K^{(-)}(\omega,z,{\bf k, q}) [-\rm{Im} ]F\sb{\sigma}({\bf q},z).
  \label{30a}
\end{eqnarray}
The kernel of the integral equations (\ref{29a}), (\ref{30a}) is determined by the relation
(see Appendix):
\begin{eqnarray}
  K^{(\pm)}(\omega,z, {\bf k, q})  =
\int\limits\sb{-\infty}\sp{+\infty} \frac{\rm d\Omega}{2\pi }  \,
 \frac{\mbox{tanh}(z/2T)+\mbox{coth}(\Omega/2T) }{\omega-z-\Omega}
\nonumber \\
 \times \big\{ |g_{sf}({\bf k, q})|^{2} {\rm Im} \chi\sb{sf}({\bf k- q},\Omega)
\nonumber \\
\pm |g_{ep}({\bf k -q})|^2 {\rm Im} \chi_{ph}({\bf k-q}, \Omega)
\nonumber \\
\pm \left[ |V({\bf k -q})|^2 + |t({\bf q})|^{2}/4 \right]
 {\rm Im}\, \chi_{cf}({\bf k-q}, \Omega)\big\}
 \nonumber \\
\equiv\int\limits\sb{-\infty}\sp{+\infty} \frac{\rm d\Omega}{2\pi }  \,
 \frac{\mbox{tanh}(z/2T)+\mbox{coth}(\Omega/2T) }{\omega-z-\Omega}
\lambda^{(\pm)}({\bf k,q}, \Omega) ,
\label{31}
\end{eqnarray}
where $|g_{sf}({\bf k, q})|^{2} = |t({\bf q}) - (1/2)J({\bf k -q})|^{2}$. The  spectral densities of bosonic
excitations are determined by the dynamic susceptibility for spin${\bf S \sb{q}} $, number
(charge) $\delta N_{\bf q} = N_{\bf q} - \langle N_{\bf q} \rangle$, and lattice (phonon) $ u\sb{\bf q} $ fluctuations
\begin{eqnarray}
\chi\sb{sf}({\bf q},\omega) & = &
  -  \langle\!\langle {\bf S\sb{q} | S\sb{-q}}
\rangle\!\rangle\sb{\omega},
\label{32a} \\
 \chi\sb{cf}({\bf q},\omega) &= &
 - \langle\!\langle \delta N\sb{\bf q} | \delta N\sb{-\bf q}
   \rangle\!\rangle\sb{\omega} ,
\label{32b}\\
\chi\sb{ph}({\bf q},\omega) &= &
 - \langle\!\langle  u\sb{\bf q} | u\sb{-\bf q}
   \rangle\!\rangle\sb{\omega} .
\label{32c}
\end{eqnarray}
They are defined by the commutator GFs~\cite{Zubarev60} for the spin ${\bf S
\sb{q}} $,  number $\delta N_{\bf q} = N_{\bf q} - \langle N_{\bf q} \rangle$, and
lattice displacement (phonon) $ u\sb{\bf q} $  operators.
At first we consider solution of these equations for the normal state.

\subsection{Normal state GF}
\label{sec:4.1}

The normal state GF  in Eq.~(\ref{17a}) can be written as
\begin{eqnarray}
&& G({\bf k},\omega)= \langle  \langle X^{0\sigma}_{\bf k} |  X^{\sigma0}_{\bf k} \rangle\rangle
= \frac{Q} {\omega -{\varepsilon}({\bf k})  - M({\bf k},\omega)},
\label{34}
\end{eqnarray}
where the normal state self-energy is given by Eqs.~(\ref{29a}), (\ref{31}).
The spectral density of electronic excitation is determined by
\begin{eqnarray}
 && A({\bf  k},\omega) = - \frac{1}{\pi Q}{\rm Im} G({ \bf k},\omega + i\epsilon )
 \nonumber \\
&& =  \frac{-M''({\bf k},\omega)/ \pi} {[\omega - {\varepsilon}({\bf k})
 - M'({\bf k},\omega)]^2 + [M''({\bf k},\omega)]^2 }.
\label{35}
\end{eqnarray}
Here we introduce the real, $ M'({\bf k},\omega)$,  and imaginary, $M''({\bf k},\omega)$, parts of the self-energy:  $ M({\bf k},\omega +i\epsilon) =   M'({\bf k},\omega) +  i M''({\bf k},\omega)$.

The renormalization parameter (\ref{17b}) for the electronic energy close to the FS, $\omega \rightarrow 0 $,  reads:
\begin{eqnarray}
  Z_{\bf k}(0)=  1 -  [\partial  M({\bf k},\omega)/ \partial \omega ]_{\omega = 0} \equiv
   1 + \lambda({\bf k}) ,
\label{36}
\end{eqnarray}
where  $\lambda({\bf k}) $ is the coupling parameter.

\subsection{Superconducting state}
\label{sec:4.2}

The superconduction gap in the SCA (\ref{17d}) is determined both by the GMFA function $\Delta_{\sigma}({\bf k})$  (\ref{26}) and the anomalous self-energy component  (\ref{30a}) and (\ref{31}).   For calculation  of  superconducting  $T_c$ we can use the linear approximation for the anomalous GF  in Eq.~(\ref{30a}):
\begin{eqnarray}
  F_{\sigma} ({\bf k},\omega)
=  Q  \frac{\phi_{\sigma}({\bf k}, \omega) }
 {Z^2_{\bf k}(\omega)\,[\omega^2 - \widetilde{\varepsilon}^2
 ({\bf k},\omega)]}  ,
 \label{37}
\end{eqnarray}
where the renormalized energy $\, \widetilde{\varepsilon} ({\bf k},\omega) = [{\varepsilon}({\bf k})+ \xi_{\bf k}(\omega) ] / Z_{\bf k}(\omega) \,$. Further we consider  the gap equation  close to the FS, $\phi_{\sigma}({\bf k}) = \phi_{\sigma}({\bf k}, \omega = 0)$:
  \begin{eqnarray}
 \phi_{\sigma}({\bf k})=
\frac{1}{N} \sum_{\bf q}
   \int\limits\sb{-\infty}\sp{+\infty} \!\!\frac{d z}{\pi}\,
   \Big[\frac{V({\bf k- q}) -J({\bf k-q})}{\exp (z/T)+  1}
 \nonumber \\
 +    K^{(-)}(0,z,{\bf k, q})\Big] [-\rm{Im}]\frac{\phi_{\sigma}({\bf q})}
 {Z^2_{\bf q}(0)\,[(z+i\epsilon)^2 - \widetilde{\varepsilon}^2
 ({\bf q}, z)]} .
 \label{38}
\end{eqnarray}
Solution of this equation we consider  in Section~\ref{sec:5.3}.

\section{Results and Discussion}
\label{sec:5}

\subsection{Model parameters}

\label{sec:5.1}

To perform numerical calculations  we should introduce susceptibility  models.  For the spin-fluctuation susceptibility  (\ref{32a})  we use the model:
\begin{eqnarray}
&&     {\rm Im}\, \chi_{sf}({\bf q},\omega+i\epsilon) =
 \chi_{s}({\bf q}) \; \chi_{s}^{''}(\omega)
\nonumber\\
 & = &  \frac {\chi_Q}{1+ \xi^2 (1+ \gamma({\bf q}))} \;  \tanh
\frac{\omega}{2T} \frac{1}{1+(\omega/\omega_{s})^2} .
 \label{r1}
\end{eqnarray}
The model describes the broad energy spectrum  of spin fluctuations $\chi_{s}^{''}(\omega)$ with the cut-off  frequency $\omega_s $   and the static susceptibility $\chi_{s}({\bf q})$ with the maximum  at the AFM wave-vector $\, {\bf Q} = (\pi,\pi)$  as observed in the paramagnetic phase.
Similar model was suggested in Ref.~\cite{Prelovsek01}.   In Ref.~\cite{Vladimirov09}  we have calculated the dynamic spin susceptibility for the $t-J$ model which frequency and wave-vector dependencies  are close to the model  (\ref{r1}). It can be used in numerical calculations  but it is more convenient to adopt the analytical model  (\ref{r1}).

 The intensity of  electron interaction with spin fluctuations is determined by the maximum of the static susceptibility $\chi_{s}({\bf q})$ at  $\, {\bf Q} = (\pi,\pi)$:
 \begin{equation}
 \chi_{Q}=
 \frac{3n }{2 \omega_{s}}\,\left \{ \frac{1}{N}
\sum_{\bf q} \frac{1} {1+\xi^2[1+\gamma({\bf q})]} \right\}^{-1}.
 \label{r1a}
\end{equation}
It is very important that  $\chi_{Q}$ is not a fitting parameter but is fixed   by the normalization condition:
  \begin{eqnarray}
 \langle {\bf S}^2_{i} \rangle & = &
  \frac{1}{\pi} \, \int\limits_{-\infty}^{+\infty }
  \frac{ \chi_{s}^{''}(\omega)\,d\omega}{\exp{(\omega/T)} - 1}\,
  \frac{1}{N} \sum_{\bf q} \chi_{s}({\bf q})
  \nonumber \\
  & = &\frac{\omega_{s}}{2} \frac{1}{N} \sum_{\bf q} \chi_{s}({\bf q}) = \frac{3n}{4},
 \label{r2}
\end{eqnarray}
that yields  $\chi_{Q} = 2 C_{\bf Q}/ \omega_s $ presented in Table~\ref{Table1}.
Therefore, there are two parameters which determine the function (63): the AFM correlation length $\xi$ that depends on the hole concentration $\delta$ as given in Table~\ref{Table1}, and the frequency $\omega_s$. The latter  can be  estimated as $\omega_s = J = 0.4\, t$  taking into account theoretical calculations, e.g., Refs.~\cite{Prelovsek01},~\cite{Vladimirov09}, and experimental results of  the inelastic magnetic neutron scattering experiments and optical measurements.
As shown in  Table~\ref{Table1}, at large correlation length $\xi$, low doping, the spin-fluctuation interaction given by $\chi_{Q}$ is strong  while with doping and decreasing $\xi$ the  interaction becomes weak.

For the EPI we can use the model of forward scattering. It can explain a weak transport EPI $\lambda_{tr}$,  while it may result in a strong superconducting coupling $\lambda$,  e.g., $\lambda_{tr} < \lambda/3$~\cite{Kulic00}. To take into account the importance of the forward scattering  we consider a model EPI  for optic phonons in Eq.~(\ref{31}) suggested in Ref.~\cite{Lichtenstein95}:
\begin{equation}
 |g_{ep}({\bf k-q})|^2 \chi_{ph}( \omega) = g_{ep} \frac{\,\xi_{ch}}{1 + \xi_{ch}^2 |{\bf k-q}|^2} \frac{\omega_0^2 }{ \omega_0^2 - \omega^2},
   \label{r3}
\end{equation}
where $\omega_{0} = 0.1\,t$ is an optic excitation frequency. The parameter $\xi_{ch}$ is the charge correlation length for holes  and can be approximated by the relation $\xi_{ch}/a = 1/ (2 \delta) $~\cite{Zeyher96}.   This results  in a large EPI in the underdoped case, while   in the  overdoping region  it  decreases, e.g., for $qa = 0.1$ it  changes from $4 g_{ep}$ at $\delta = 0.1 $  to $1.62 g_{ep}$ at $\delta = 0.3$.   We assume  a large EPI coupling constant  $g_{ep} = 8\,t = 3.2$~eV.

For the charge  susceptibility  (\ref{32b}) we use the model considered in our calculation of the charge density waves in Ref.~\cite{Plakida18}:
\begin{eqnarray}
 \chi_{cf}({\bf q},\omega) & = &\frac{4}{\Omega_{\bf q}^2 -\omega^2}\frac{1}{N}\sum_{\bf q'}\, [\, t({\bf q}')
 - t({\bf q' - q})]\, N({\bf q}),
 \nonumber \\
 \Omega_{\bf q}^{2}& = &\frac{2}{N} \sum _{{\bf q}' }
 \left[ t ({\bf q}') -t ( {\bf q} - {\bf q}')  \right]
\label{r4}\\
  & \times & \left[  t ( {\bf q}') - (1/2)\, J({\bf q}) + 2 V({\bf q}) \right ]\, N({\bf q}) ,
  \nonumber
\end{eqnarray}
where the electronic occupation number $N({\bf q})$ we calculate in the GMFA  (\ref{20}).

We perform numerical calculations  for the same  parameters  as for the electronic spectrum in Section~\ref{sec:3.1} and given in the Table~\ref{Table1}.  The calculations are done for low temperature $T= 0.02$  which is much less than the chemical potential  and the exchange interaction $J = 0.4$  so we can neglect temperature dependence of the correlation length and the electronic spectrum.

\subsection{Normal state}
\label{sec:5.2}

Electronic spectrum in the normal state is determined by the the spectral density  (\ref{35}). To calculate it we should find the  self-energy (\ref{29a}) given by Eq.~(\ref{31}):
\begin{eqnarray}
&&  M({\bf k},\omega) =\frac{1}{2\pi N} \sum\sb{\bf q}
   \int\limits\sb{-\infty}\sp{+\infty}\!\!{}\,
   \int\limits\sb{-\infty}\sp{+\infty} \frac{ d z \, d\Omega} {\omega - z- \Omega} \,
 \nonumber\\
&& \times \, \Big[\tanh\frac{z}{2T}+\coth\frac{\Omega}{2T} \Big]\lambda^{+}({\bf k,q}, \Omega) A({\bf  q},z).
   \label{29b}
\end{eqnarray}
We calculate the self-energy and  the spectral density by iteration. In the lowest order for  the spectral density  $A^{(0)}({\bf  q},z) = \delta (z - {\varepsilon}({\bf q}))$  the first order of  the imaginary part of the self-energy  reads
\begin{eqnarray}
&&-\frac{1}{\pi}{\rm Im}  M^{(1)}({\bf k},\omega + i\epsilon)
= \frac{1}{2 \pi N} \sum\sb{\bf q} \lambda^{+}({\bf k,q}, \omega-{\varepsilon}({\bf q}))
\nonumber\\
&& \times \, \Big[\tanh\frac{{\varepsilon}({\bf q})}{2T}
+\coth\frac{\omega-{\varepsilon}({\bf q})}{2T} \Big] .
   \label{50c}
\end{eqnarray}
The real part of the self-energy (\ref{29b}) is calculated using the dispersion relation for the GFs~\cite{Zubarev60}. The   $n$-order of the self-energy $ M^{(n)}({\bf k},\omega)$ is calculated  using the $(n-1)$-order of the spectral density $A^{(n-1)}({\bf  q}, \omega)$.
The iteration procedure converges for $n > 8$ and we present the results of calculations in the $n=10$-order of iterations.
For  the interactions in Eq.~(\ref{31}) we calculate separately contributions to  the imaginary part of the self-energy  determined by  spin fluctuations $M_{sf}({\bf  k},\omega)$, phonons $M_{ph}({\bf  k},\omega)$ and charge fluctuations (CF) $M_{cf}({\bf  k},\omega)$.

Let us consider the contribution to the self-energy  (\ref{29b}) produced by spin fluctuations.
For the model  (\ref{r1})  the first order of the imaginary part (\ref{50c}) is given by
\begin{eqnarray}
- \frac{1}{\pi}\,{\rm Im}  M_{sf}^{(1)}({\bf k},\omega)  =
  \frac{1}{2 \pi N} \sum_{\bf  q}\frac {|g_{sf}({\bf k,  q})|^{2}\,\chi_Q}
   {1+ \xi^2 (1+ \gamma({\bf  k-q}))}
\nonumber\\
 \times \frac{ \tanh (\varepsilon({\bf q})/2T)\,\tanh [(\omega- {\varepsilon}({\bf q}) ) /2T]  +1 } {1+[(\omega - {\varepsilon}({\bf q}))/\omega_{s}]^2} .
\label{42a}
\end{eqnarray}
As we show later, contributions to the  imaginary part of the self-energy from phohons  and CF are much smaller than from spin-fluctuations. Therefore, we can neglect these contributions and in the iteration procedure for the self-energy and the spectral density  use $ M_{sf}({\bf k},\omega)$ and $ A_{sf}({\bf  k},\omega)$.

\begin{figure}[htp]
  \centering
  \includegraphics [scale=0.6]{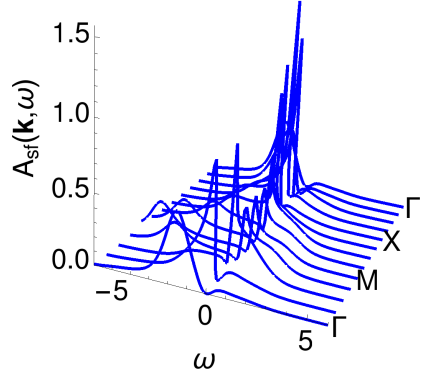}
  \caption{ Spectral density $A({\bf  k},\omega)$   along the main
directions in the BZ: $\Gamma (0, 0) \rightarrow M(\pi, \pi)
\rightarrow X(\pi, 0) \rightarrow  \Gamma(0, 0)$ for $\delta = 0.05$. }
\label{fig5}
\end{figure}

\begin{figure}[htp]
  \centering
  \includegraphics [scale=0.6]{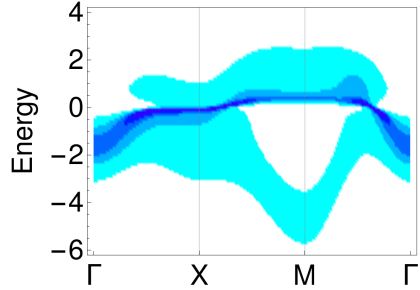}
  \caption{ Energy dispersion  along the main
directions in the BZ: $\Gamma \left ( 0,0 \right ) \rightarrow X\left ( \pi,0 \right )
\rightarrow M\left ( \pi,\pi \right ) \rightarrow
 \Gamma \left ( 0,0 \right )$ for $\delta = 0.05$. }
\label{fig6}
\end{figure}

\begin{figure}[htp]
  \centering
  \includegraphics [scale=0.6]{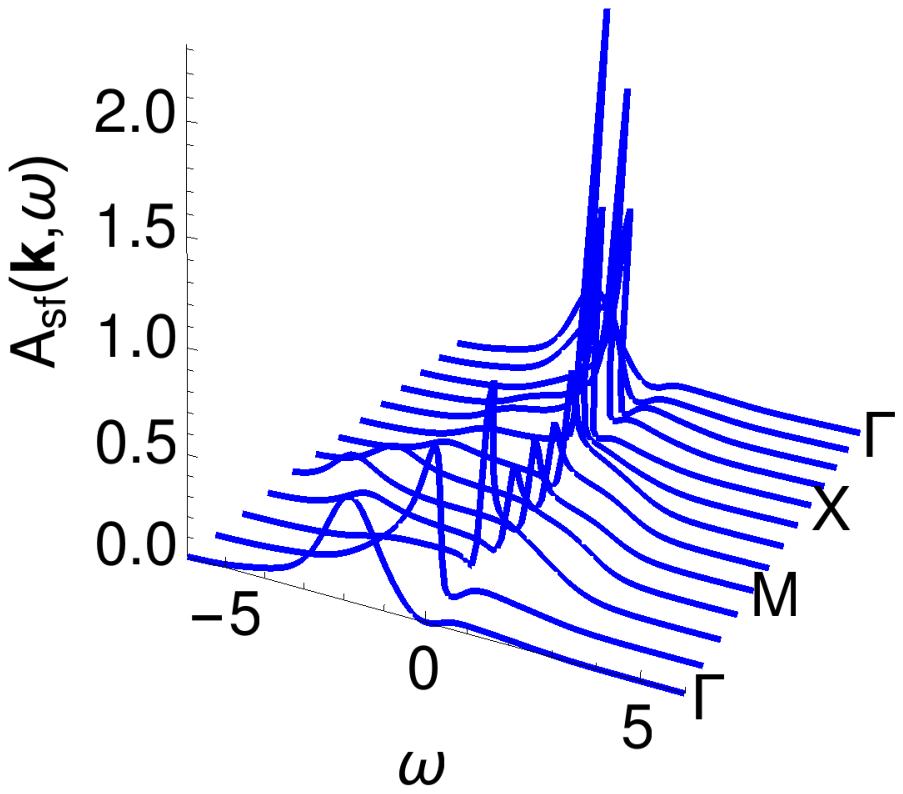}
  \caption{  Spectral density $A({\bf  k},\omega)$  for $\delta = 0.1$. }
\label{fig7}
\end{figure}

\begin{figure}[htp]
  \centering
  \includegraphics [scale=0.6]{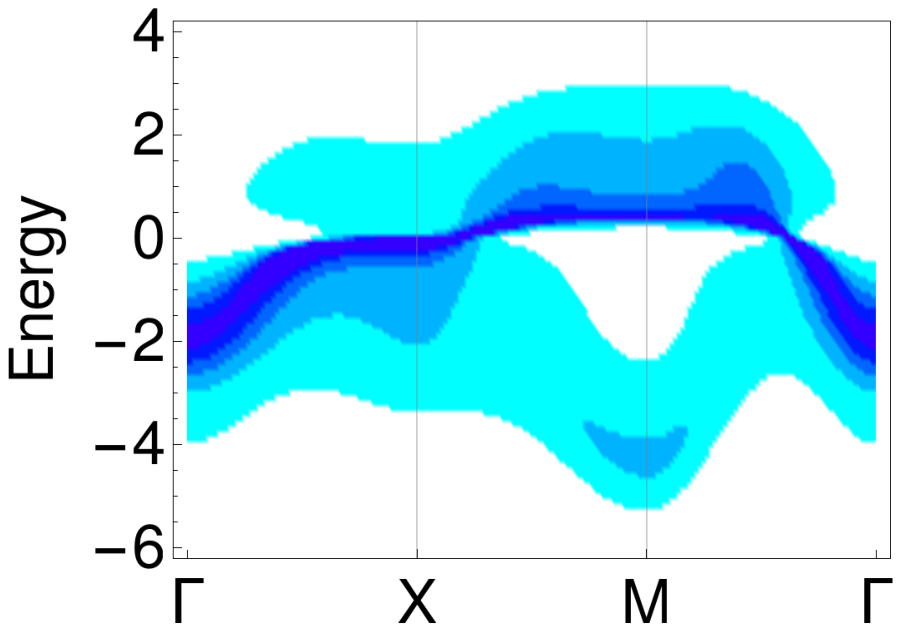}
  \caption{ Energy dispersion  for $\delta = 0.1$}.
\label{fig8}
\end{figure}

\begin{figure}[htp]
  \centering
  \includegraphics [scale=0.6]{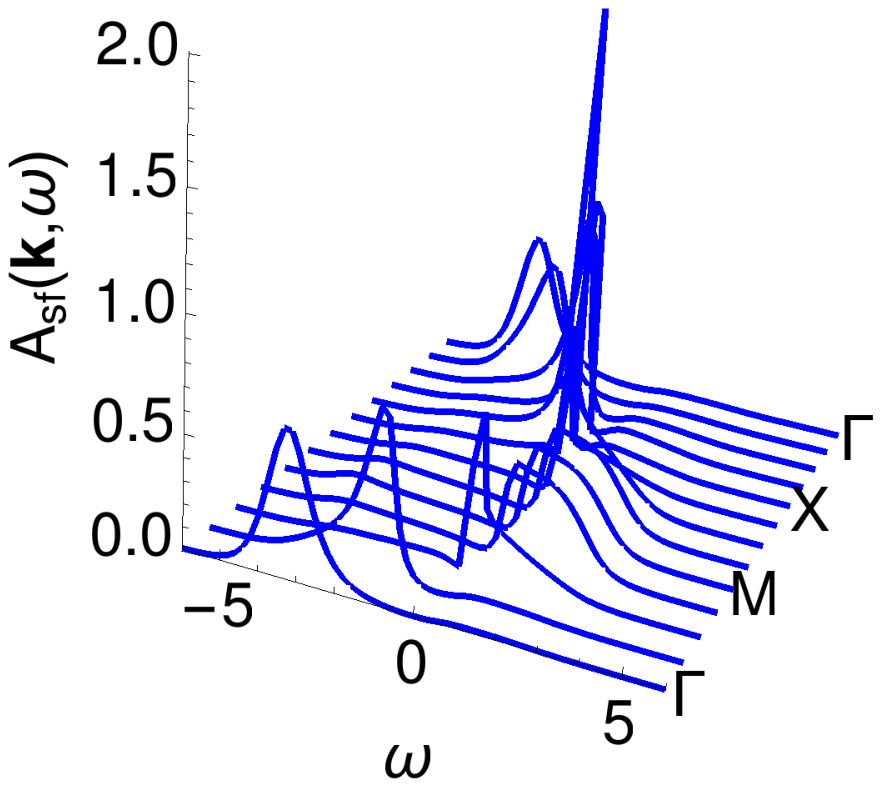}
  \caption{ Spectral density $A({\bf  k},\omega)$    for $\delta = 0.3$. }
\label{fig9}
\end{figure}

\begin{figure}[htp]
  \centering
  \includegraphics [scale=0.6]{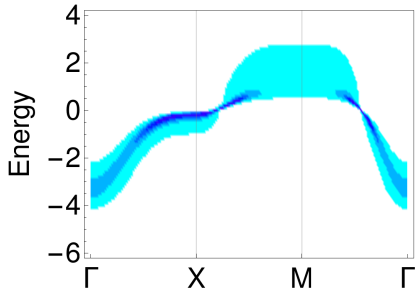}
  \caption{ Energy dispersion  for  $\delta = 0.3$.}
\label{fig10}
\end{figure}

The results of the 10-th order of iterations for the  spectral density $A_{sf}({\bf  k},\omega)$ (\ref{35}) and the energy dispersion $\tilde{\varepsilon}({\bf  k})$ obtained by 2D projection of $A_{sf}({\bf  k},\omega)$ along the main
directions in the BZ, $\Gamma \left ( 0,0 \right ) \rightarrow X\left ( \pi,0 \right )
\rightarrow M\left ( \pi,\pi \right ) \rightarrow
 \Gamma \left ( 0,0 \right )$,  are presented in Figs.~\ref{fig5} -- \ref{fig10}.
At low doping the spectral density shows a large incoherent background, in particular close  to the  $(\pi,\pi)$-point of the BZ, as shown in Figs.~\ref{fig5}, \ref{fig6} for $\delta = 0.05$ and  in Figs.~\ref{fig7}, \ref{fig8} for $\delta = 0.1$.  With increasing doping the spin-fluctuation interaction becomes weak and  the incoherent background decreases, while  the intensity of excitations increases as shown in Figs.~\ref{fig9}, \ref{fig10} for $\delta = 0.3$. The spectrum of excitations at large doping in Fig.~\ref{fig10} is close to that one in the GMFA  shown in Fig.~\ref{fig1}. However,  at low doping where the self-energy renormalization is  strong the spectrum in the GMFA  is quite different from those shown in Figs.~\ref{fig6},~\ref{fig8}. In particular, a large intensity of excitations  at  the $(\pi,\pi)$-point of the BZ in Fig.~\ref{fig6} appears at much  lower energy than in  the GMFA due to a  shift of the excitation energy caused by the real part of the self-energy. Therefore, we can conclude that the self-energy effects are very important in studies of the QP excitations in the  $t$-$J$ model.

\begin{figure}[htp]
\centering
\includegraphics [scale=0.6]{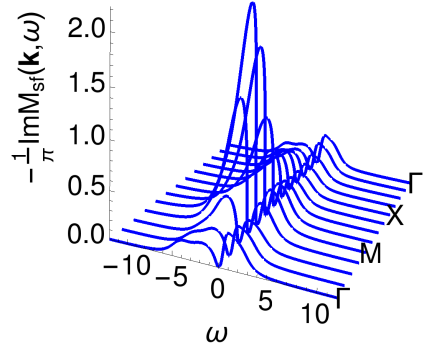}
\caption{ Imaginary part of the spin-fluctuation self-energy   $- (1/\pi){\rm Im} M_{sf}({\bf k},\omega)$ for $\delta = 0.1$.}
\label{fig11}
\end{figure}

\begin{figure}[htp]
\centering
\includegraphics [scale=0.6]{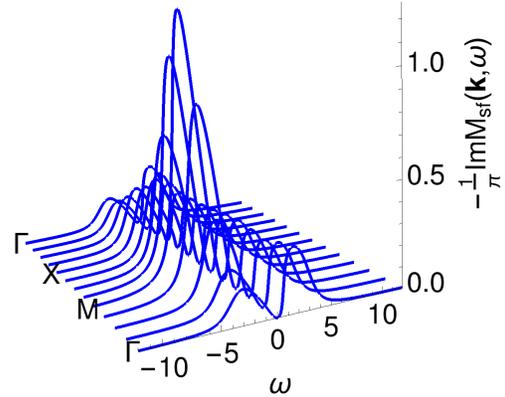}
\caption{ Imaginary part of the spin-fluctuation self-energy   $- (1/\pi){\rm Im} M_{sf}({\bf k},\omega)$ for $\delta = 0.3$.}
\label{fig12}
\end{figure}

The QP  damping   determined by the imaginary  part of the self-energy (\ref{29b}) $\Gamma({\bf k},\omega) =- (1/\pi){\rm Im} M_{sf}({\bf k},\omega)$ due to spin-fluctuation interaction is plotted in Fig.~\ref{fig11} at doping $\delta = 0.1$.  For a larger doping, $\delta = 0.3$, the intensity decreases as shown in Fig.~\ref{fig12}. A large asymmetry of the damping for the hole spectrum below the Fermi energy,  $\omega < 0$, and for the electron spectrum at  $\omega >  0$ is observed  with a strong  damping for the  hole spectrum.  In Fig.~\ref{fig12} for $\delta = 0.3$  it is shown more  clearly.
Strong damping for the hole spectrum  results in a large incoherent background in the  spectral density $A({\bf  k},\omega)$   in Figs.~\ref{fig5} -- \ref{fig10}. Close to  the Fermi energy the damping  $\Gamma({\bf k},\omega \rightarrow 0)$ disappears linearly with $\omega$  for a small doping as in  Fig.~\ref{fig11}, while for a large doping it proportional to $\omega^2$ as in  Fig.~\ref{fig12}. This behavior looks like as a transition from the marginal Fermi-liquid to the conventional Fermi-liquid.

In comparison with other studies we can mention the spectrum of excitations for the $t$-$J$ model    obtained within the CPT in Ref.~\cite{Kohno15} (see Fig.1 (b-d)). The spectrum  is close to our results, in particular, a  flat  dispersion for the  energy excitations with high intensity close  $X (\pi,0)$ point of the BZ is clearly reproduced in our figures.
Studies of the spectral properties for  the two-dimensional $t$-$J$ model  using the finite-temperature Lanczos method in Ref.~\cite{Jaklic97} have revealed similar results for  the spectral density and the imaginary part of the self-energy. The spectrum of the hole excitations found in the   Hubbard model reproduces the main features of the spectrum in the $t$-$J$ model as was found within the equation of motion method  in Refs.~\cite{Plakida07},  \cite{Plakida13} and  applying  the CPT in  Ref.~\cite{Kuzmin20}.

\begin{figure}[htp]
\centering
\includegraphics [scale=0.6]{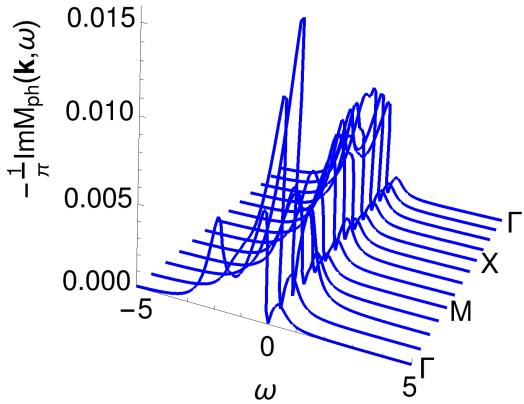}
\caption{Imaginary part of the phonon self-energy   $- (1/\pi){\rm Im} M_{ph}({\bf k},\omega) $ for $\delta = 0.1$.}
\label{fig13}
\end{figure}

The EPI  contribution  (\ref{r3}) for  the imaginary part of the phonon self-energy $M_{ph}({\bf k},\omega)$ (\ref{50c}) reads:
\begin{eqnarray}
-\frac{1}{\pi}{\rm Im} M_{ph}({\bf k},\omega)
= \frac{1}{2 \pi N} \sum\sb{\bf q}  \frac{g_{ep} \xi_{ch}}{1  + \xi_{ch}^2 |{\bf k-q}|^2 }
 \nonumber\\
 \times\int\limits\sb{-\infty}\sp{+\infty} d z \, \Big[\tanh\frac{z}{2T}+\coth\frac{\omega-z}{2T} \Big]
 {\rm Im} \frac{A({\bf  q},z)\,  \omega_0^2}{ \omega_0^2 -[\omega- z]^2 }.
  \label{42b}
\end{eqnarray}
The frequency dependence of the imaginary part of the phonon self-energy   is shown in Fig.~\ref{fig13} for $\delta = 0.1$. For the spectral density we used  $A_{sf}({\bf  q},z)$.
\begin{figure}[htp]
\centering
\includegraphics [scale=0.5]{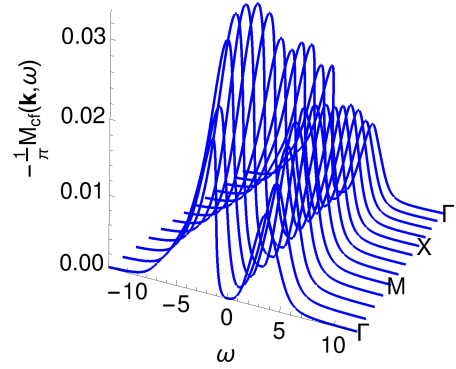}
\caption{Imaginary part of the charge fluctuation self-energy   $- (1/\pi){\rm Im} M_{sc}({\bf k},\omega) $ for $\delta = 0.1$.}
\label{fig14}
\end{figure}
The CF contribution  (\ref{r4}) for  the imaginary part  of the self-energy $M_{cf}({\bf k},\omega)$ (\ref{50c}) is given by
\begin{eqnarray}
 -\frac{1}{\pi}{\rm Im} M_{cf}({\bf k},\omega)= \frac{1}{2 \pi N} \sum\sb{\bf q}
  \left[ |V({\bf k -q})|^2 + \frac{1}{4}|t({\bf q})|^{2} \right]
  \nonumber\\
   \times \!\!\! \int\limits\sb{-\infty}\sp{+\infty} d z \Big[\tanh\frac{z}{2T}+\coth\frac{\omega-z}{2T} \Big]  A({\bf  q},z) {\rm Im}\, \chi_{cf}({\bf k-q},  \omega- z).
  \label{42c}
 \end{eqnarray}
The imaginary part of the CF self-energy  is plotted in Fig.~\ref{fig14} for $\delta = 0.1$.  For the spectral density we used  $A_{sf}({\bf  q},z)$.

The phonon and CF  contributions are an order of magnitude smaller than  the imaginary  part of the spin-fluctuation self-energy and, therefore, can be ignored in calculation of the spectral density. Therefore, the results obtained for the spin-fluctuation spectral density $A_{sf}({\bf  k},\omega)$ can be considered as the total spectral density $A({\bf  k},\omega)$  (\ref{35}).
\begin{figure}[htp]
\centering
\includegraphics[scale=0.6]{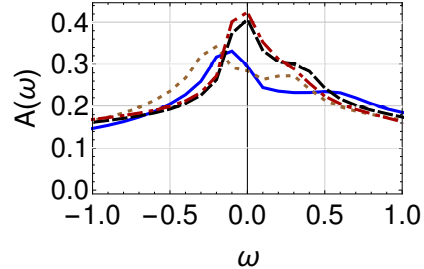}
\caption{(Color online)  Density of states $A(\omega)$ for $\delta = 0.05$ (red, dash-dotted line)) $\delta = 0.1$ (black, dashed line)), $\delta = 0.2$  (brown, dotted line), $\delta = 0.3$ (blue, solid line).}
\label{fig15}
\end{figure}
The  DOS  $A(\omega)$ in the SCA is determined by the function
\begin{equation}
A(\omega) =  \frac{1}{N } \sum_{\bf k} \frac{1}{\pi Q} [- {\rm Im} G({\bf k},\omega)]=
\frac{1}{N } \sum_{\bf k} A({\bf  k},\omega)  ,
\label{22a}
\end{equation}
and presented in Fig.~\ref{fig15} in units of $1/t$ for various doping.

In comparison with the  DOS  $A^{(0)}(\omega)$ in the GMFA  in  Fig.~\ref{fig3},   $A(\omega)$  in the SCA shows lower values for small doping at the FS due to a small QP weight $1/Z_{\bf k}$ at low doping.
\begin{figure}[h!]
\centering
\includegraphics [scale=0.5]{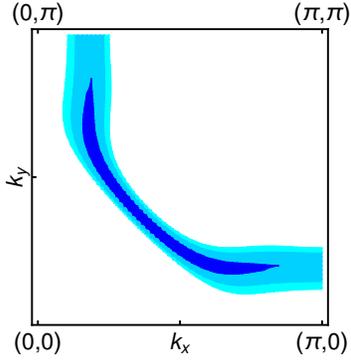}
\caption{Spectral density $A({\bf  k},  0)$ in the quarter of the BZ for $\delta = 0.05$. }
\label{fig16}
\end{figure}
\begin{figure}[h!]
\centering
\includegraphics [scale=0.5]{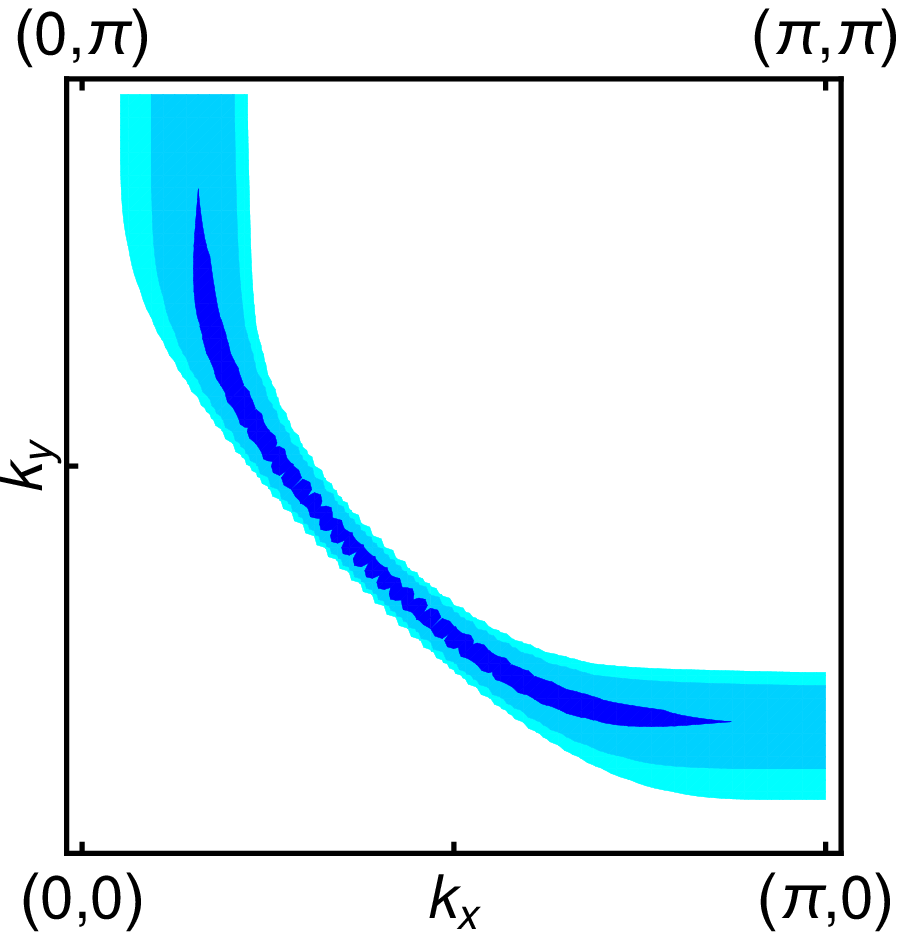}
\caption{Spectral density $A({\bf  k}, 0)$  for $\delta = 0.1$.}
\label{fig17}
\end{figure}
\begin{figure}[h!]
\centering
\includegraphics [scale=0.5]{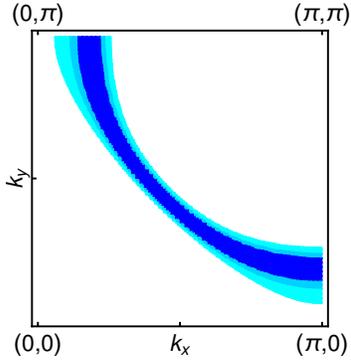}
\caption{Spectral density $A({\bf  k}, 0)$  for $\delta = 0.3$.}
\label{fig18}
\end{figure}
The results of  spectral density close to the Fermi surface $A({\bf k},\omega =0 )$  (\ref{35}) which determines the FS  are presented in Figs.~\ref{fig16} -- \ref{fig18}.  The FS changes from the arc-type at low doping ($\delta = 0.05, 0.1$ ) as demonstarted in Figs.~\ref{fig16}, \ref{fig17} to the large FS at high doping ($\delta = 0.3$) in Fig.~\ref{fig18}. In the GMFA, we have well-defined qusiparticles with the FS in the form of  hole pockets at $\delta = 0.05, \; 0.1$ which changes to large FS at large doping   as shown   in Fig.~2. Taking into account the self-energy effects,  the spectral density  $A({\bf  k},  0)$ reveals  transfer of pockets to arcs as demonstarted in Figs.~\ref{fig16}, \ref{fig17} where only the outer part of pockets is visible, while the inner part,  closer to the $(\pi,\pi)$ point of the BZ, due to a large imaginary part of the self-energy, is smoothed away.  The spectral density  for low doping also describes the pseudogap formation with  zero density of states  at around  $(\pm\pi,0), \,(0, \pm\pi)$ points of BZ. Similar behavior for the spectral density at the FS was found in Ref.~\cite{Tohyama04} in the exact diagonalization studies for the $t$-$t'$-$t''$-$J$ model. At low hole concentration $\delta = 0.1$  a gap opening occurs in this region  leading to  arcs on the FS. The arc type FS  were obtained using the CPT for the $t$-$J$ model in Ref.~\cite{Kohno15} and for the  Hubbard model in  Ref.~\cite{Kuzmin20}.

In ARPES experiments in cuprates only  the arc transformation at low doping to large FS is observed since a weak intensity of the inner part of the pockets makes them invisible. In particular,
in Ref.~\cite{Shen05}, the FS show the arcs at doping $x = 0.05,\;  0.1$ which can be considered as a manifestation of  the pockets where only the outer part of it is found.   Similar results were obtained in other publications, see  Ref.~\cite{Kordyuk02} where large intensity of the ARPES signal at the  "arc-type" part of the FS in the underdoped regime and the large FS in overdoped regime were obtained. In Ref.~\cite{Lee07} two gaps were found, one is the PG region  outside the arcs and the SC gap related to arcs.  In Ref.~\cite{Hashimoto08} well defined arcs on the FS with underlying FS are obtained.

\begin{figure}[ht]
\centering
\includegraphics [scale=0.6]{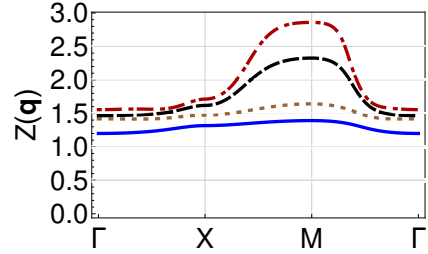}
 \caption{(Color online) Doping dependence of renormalization  parameter $Z(\bf{q})$ for $\delta = 0.05$ (red, dash-dotted line)) $\delta = 0.1$ (black, dashed line)), $\delta = 0.2$  (brown, dotted line), $\delta = 0.3$ (blue, solid line). }
\label{fig19}
\end{figure}

The wave-vector and doping dependence  of the QP renormalization parameter
$ Z_{\bf k}$~(\ref{36}) calculated from the real part of the self-energy  is shown in Fig.~\ref{fig19}. It strongly depends on the doping being especially large for low doping resulting in a small QP weight $1/Z_{\bf k}$.

\subsection{Superconducting state}

\label{sec:5.3}

The gap \, equation (\ref{38}) close to the FS defined by the function $ K^{(-)}(0,z, {\bf k, q}) $
(\ref{31}) can be written as
\begin{eqnarray}
  \phi_{\sigma}({\bf k})&=&
\frac{1}{N } \sum_{{\bf q}}[J({\bf k-q}) - V({\bf k- q})]\frac{\phi_{\sigma}({\bf q})}{2 Z^2_{\bf q} \widetilde{\varepsilon}({\bf q})} \tanh\frac{ \widetilde{\varepsilon}({\bf q}))}{2T_c}
\nonumber \\
& - &\frac{1}{N} \sum_{\bf q}  \int\limits\sb{-\infty}\sp{+\infty}
  \int\limits\sb{-\infty}\sp{+\infty} \frac{dz\, d\Omega}{2\pi }  \,
 \frac{\tanh(z/2T)+\coth(\Omega/2T) }{ z+\Omega}
\nonumber \\
 &\times&\, \lambda^{(-)}({\bf k,q}, \Omega)\,
 [-\mbox{Im}]\frac{\phi_{\sigma}({\bf q})/\pi}
 {Z^2_{\bf q}\,[z^2 - \widetilde{\varepsilon}^2 ({\bf q})]},
\label{sc1}
\end{eqnarray}
where the first term is the integrated over $z$ in  (\ref{38})  the  GMFA gap function. The second order contributions are determined by the function:
\begin{eqnarray}
&& \lambda^{(-)}({\bf k,q}, \Omega) =
\big\{ |g_{sf}({\bf k, q})|^{2} {\rm Im} \chi\sb{sf}({\bf k- q},\Omega)
\nonumber \\
&&- |g_{ep}({\bf k -q})|^2 {\rm Im} \chi_{ph}({\bf k-q}, \Omega)
\nonumber \\
&& - \left[ |V({\bf k -q})|^2 + |t({\bf q})|^{2}/4 \right]
 {\rm Im}\, \chi_{cf}({\bf k-q}, \Omega)\big\}
 \nonumber \\
 &&\equiv  -\sum_{\alpha}\,|g_{\alpha}({\bf k, q})|^2  {\rm Im} \chi_{\alpha}({\bf k-q}, \Omega),
\label{sc2}
\end{eqnarray}
where $\alpha = sf, ph, cf$. After integration  over $z$ of the second term  in (\ref{sc1}) for  the  $\alpha$-component of this term we obtain:
\begin{eqnarray}
\phi^\alpha_{\sigma}({\bf k})  =
 \frac{1}{N} \sum_{\bf q}  \frac{\phi_{\sigma}({\bf q})|g_{\alpha}({\bf k, q})|^2}{2 Z^2_{\bf q} \widetilde{\varepsilon}({\bf q}) } \int\limits\sb{-\infty}\sp{+\infty} \frac{d\Omega}{2\pi }
 \nonumber \\
 \times  \frac{ {\rm Im} \chi_{\alpha}({\bf k-q}, \Omega)}{ \Omega^2 - \widetilde{\varepsilon}^2({\bf q})}
\Big[2\Omega \tanh \frac{\widetilde{\varepsilon}({\bf q})}{2T} - 2\widetilde{\varepsilon}({\bf q})\coth \frac{\Omega}{2T}   \Big] .
 \label{sc3}
\end{eqnarray}
Here summation over ${\bf q}$  is performed  for the electronic energy $\widetilde{\varepsilon}({\bf q})$ close to the Fermi energy $\widetilde{\varepsilon}({\bf q}) = 0$  in the narrow region $|\widetilde{\varepsilon}({\bf q})| < \Omega  $. The bosonic excitation energy $\Omega$ is determined by the dynamic susceptibility $\, {\rm Im} \chi_{\alpha}({\bf k-q}, \Omega)\,$. In this approach we can take  the electronic energy $\widetilde{\varepsilon}({\bf q}) = 0$  in  the denominator $ \Omega^2 - \widetilde{\varepsilon}^2({\bf q})$  of Eq.~(\ref{sc3})   and neglect the contribution from the bosonic excitations given by  $ \, \widetilde{\varepsilon}({\bf q})\coth ({\Omega}/{2T})$.  In this approximation  Eq.~(\ref{sc3}) reads:
 \begin{eqnarray}
\phi^\alpha_{\sigma}({\bf k}) =
  \frac{1}{N} \sum_{\bf q}  \,|g_{\alpha}({\bf k, q})|^2
 \chi_{\alpha}({\bf k-q})
 \frac{\phi_{\sigma}({\bf q})}{2 Z^2_{\bf q} \widetilde{\varepsilon}({\bf q}) }
 \tanh \frac{\widetilde{\varepsilon}({\bf q})}{2T} ,
 \label{sc4}
\end{eqnarray}
where we took  into account the dispersion relation for the susceptibility
$\int_{-\infty}^{+\infty}({d\Omega}/{\pi \Omega } ) {\rm Im} \chi_{\alpha}({\bf k-q}, \Omega)
={\rm Re}\chi_{\alpha}({\bf k-q}, 0)$  and introduced the static susceptibility $\chi_{\alpha}({\bf k-q}) ={\rm Re}\chi_{\alpha}({\bf k-q}, 0) $.  Therefore, the gap equation (\ref{sc1})  for $\phi_{\sigma}({\bf k}) ={\sigma} \phi({\bf k}) $ takes the form
\begin{eqnarray}
&&  \phi({\bf k})=
\frac{1}{N } \sum_{{\bf q}}\frac{\phi({\bf q})}{2 Z^2_{\bf q}\, \widetilde{\varepsilon}({\bf q})} \tanh\frac{ \widetilde{\varepsilon}({\bf q}))}{2T_c}
\nonumber \\
&&\times\Big\{ J({\bf k-q}) - V({\bf k- q})
\label{sc5} \\
&- &|t({\bf q})|^{2}  \chi\sb{sf}({\bf k- q}) \theta(\omega_{s} -
   |\widetilde{\varepsilon}({\bf q})|)
\nonumber \\
&+& |g_{ep}({\bf k -q})|^2  \chi_{ph}({\bf k-q}) \theta(\omega_{0} -
   |\widetilde{\varepsilon}({\bf q})|)
\nonumber \\
&+& \Big[ |V({\bf k -q})|^2 + \frac{1}{4} |t({\bf q})|^{2} \Big]
 \chi_{cf}({\bf k-q}) \theta(\omega_{c} -
   |\widetilde{\varepsilon}({\bf q})|)\Big\}, \nonumber
\end{eqnarray}
where   $\theta$-functions  restrict the integration over ${\bf q}$  for $|\widetilde{\varepsilon}({\bf q})| < \Omega$. To compare contributions for pairing from the spin fluctuations and the EPI we take into account in the interaction $\,|g_{sf}({\bf k, q})|^2\,$  only the first term $|t({\bf q})|^{2} $.
\begin{figure}
\centering
\includegraphics[scale=0.4]{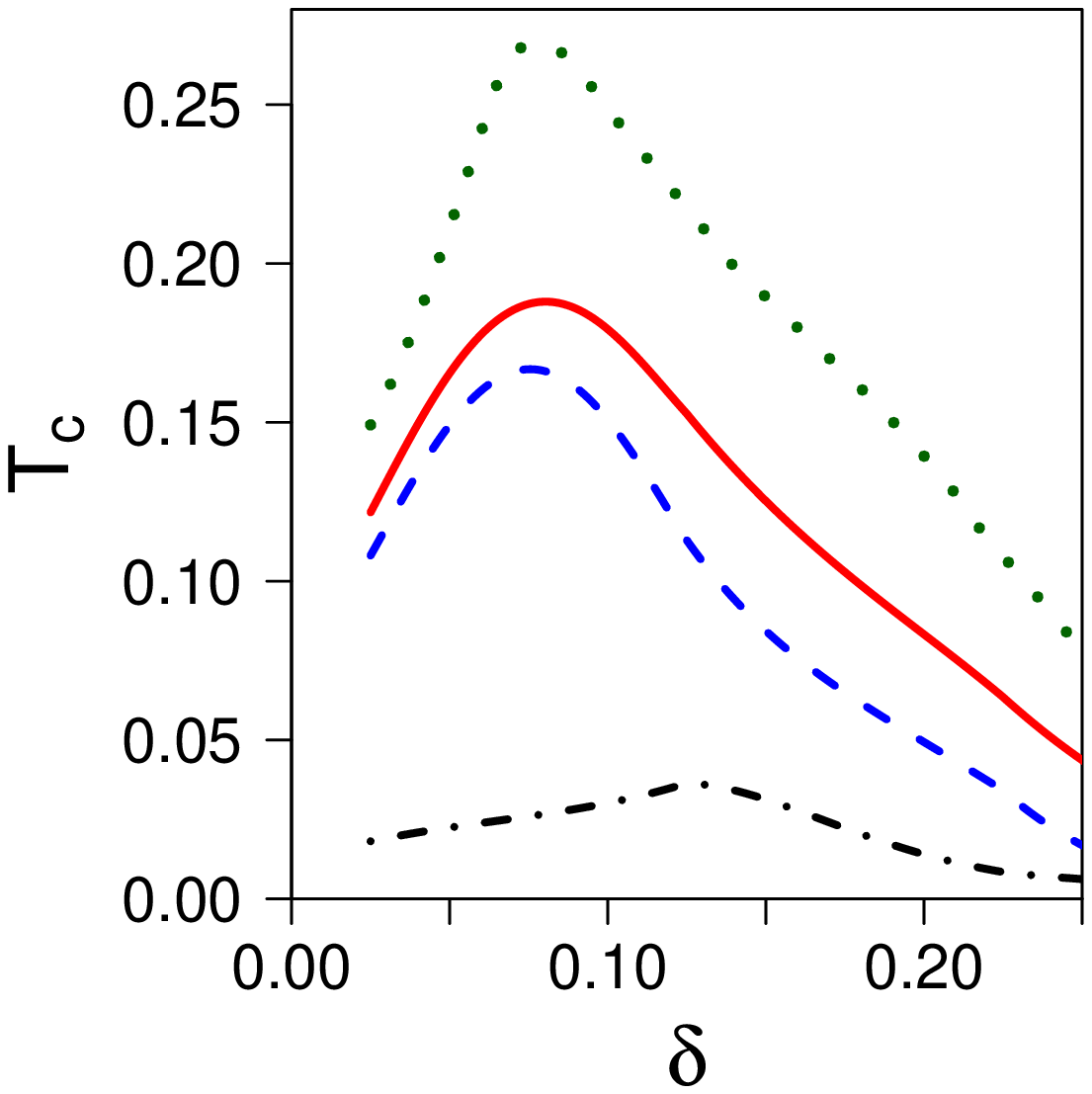}
\caption{(Color online) Solution of the gap equation (\ref{sc5}) in the WCA,  $Z =1$, for $T^{ep}_c$ (black, dash-dotted ed line), $T^{sf}_c$ (blue, dashed line), and  $T^{sf+ep}_c$ (red, solid line). The green dotted line show $T^{sf+ep}_c$ with zero CI, $V({\bf k -q}) = 0$.}
 \label{fig20}
\end{figure}
\begin{figure}
 \centering
\includegraphics[scale=0.4]{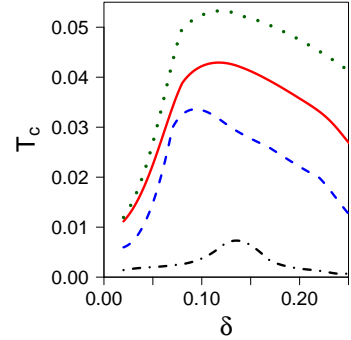}
\caption{(Color online) Solution of the gap equation (\ref{sc5}) in the SCA for $T^{ep}_c$ (black, dash-dotted line), $T^{sf}_c$ (blue, dashed line), and  $T^{sf+ep}_c$ (red, solid line). The green dotted line show $T^{sf+ep}_c$ with zero CI, $V({\bf k -q}) = 0$.}
 \label{fig21}
 \end{figure}
Solution of the gap equation (\ref{sc5}) in the WCA, $Z({\bf q}) = 1$, for    $T_c$ as a function of doping is presented in  Fig.~\ref{fig20}. Solution of the  gap equation (\ref{sc5}) for $T_c$ in the SCA for $Z(\bf{q}) $ given in Fig.~\ref{fig19} is shown in   Fig.~\ref{fig21}. To simplify the numerical calculation  we approximated the function $Z(\bf{q}) $  by its  average over $\bf{q}$ values: $ Z = 2.1, 1.7, 1.3$ for $\delta = 0.1, 0.2, 0.3$,  respectively, which can be described by the function  $Z = 2.5 - 4\delta$.
The superconducting  $T_c$ in the SCA in Fig.~\ref{fig21} is an order of magnitude smaller than in the WCA in Fig.~\ref{fig20} due to suppression of the QP weight given by $1/Z(\bf{q}) $.

To emphasize existences of  two channels of pairing in our theory, induced by the unretarded interaction $J$ and the retarded spin-fluctuation interaction, we include in Figs.~\ref{fig20} and \ref{fig21} the results for $T_c$ without CI $V$ (dotted green line). In this case there is no suppression of the interaction $J$ by CI $V$, as in the GMFA in  Fig.~\ref{fig4}, and  both contributions to $T_c$ are revealed. We see that the interaction $J$ considerably  enhance  $T_c$ but the main contribution comes from the retarded  EPI  and interaction with spin-fluctuations.  A more detailed  discussion of the role of CI $V$ was given for the Hubbard model in Ref.~\cite{Plakida14}. It was shown that spin-fluctuations is a separate channel of pairing, which can be suppressed only for CI  $V \gtrsim 3 t $,  much larger than the exchange interaction  $J = 0.4 t$. The same result holds for the $t-J-V$ model not presented  in the  manuscript for shortness reason.
Existence of  two components of  pairing interaction induced by the nonretarded exchange interaction $J$ and retarded spin-fluctuation interaction  in the Hubbard and $t$--$J$ models was discussed in Ref.~\cite{Maier08}. It was concluded that the basic is the retarded spin-fluctuation interaction which  can be considered as a ``glue'' which mediates the $d$-wave pairing. This conclusion was confirmed in ARPES experiments, see, e.g., Ref.~\cite{Kordyuk04}.

The role of intersite CI is also discussed  in Ref.~\cite{Plekhanov03}.
 Using the variational Monte Carlo technique the
superconducting $d$-wave  gap was calculated for  the extended Hubbard model with a weak
exchange interaction $J = 0.2 \, t$  and a repulsion $V \leq 3\, t $ in a broad range of
$\,0 \leq U \leq 32$. It was found that the gap decreases with increasing $V$ at all $U$
and can be suppressed for  $V > J $ for small $U $. But for large $ \,U \gtrsim U_c \sim
6\, t\,$ the gap becomes  robust and exists up to  large values of $V \sim 10\, J = 2\, t
$. At the same time, the gap does not show notable variation with $U$ for large $U = 10 - 30$
though it should depend  on the conventional exchange interaction in the Hubbard model $J
= 4t^2/U$.  We can  explain   these results by pointing out that at large $U \gtrsim U_c \,$ concomitant decrease of the bandwidth  in Ref.~\cite{Plekhanov03}) results in the splitting of the Hubbard band into the upper and lower subbands and  emerging the kinematic interaction
which induces the $d$-wave pairing in one Hubbard subband. In that case the second subband for
large $U$ gives a small contribution which results in $U$-independent pairing. It can be
suppressed by the repulsion $V$ only larger than the kinematic interaction, $V \gtrsim
2t$.

We calculated also  $T_c$ for several next-nearest-neighbor  parameters $t'$. It was found that $T_c^{max}$  increases with increasing of  $t'$ as was found also in Ref.~\cite{Prelovsek05}. Variation of $t'$ results in changing of the FS but it also changes the spin-fluctuation interaction $|t({\bf q})|^{2} $. For the FS close to the AFM BZ,  $ X(\pi,0)\rightarrow Y(0,\pi)$,  ($ \cos k_x + \cos k_y = 0$), the interaction given by the hopping parameter  $ 2t \, (\cos k_x + \cos k_y )$  is weak while the interaction determined  by the  hopping parameter $t' \cos k_x  \cos k_y $ gives a substantial  contribution for the $d$-wave symmetry order parameter $\phi_{\sigma}({\bf q}) \propto \cos k_x -\cos k_y $.  Increasing of $T_c^{max}$ with $t'$  was found in the band-structure calculations ~\cite{Pavarini01} and observed in ARPES experiments~\cite{Tanaka04}.

Comparison of $T_c$ in the WCA in Fig.~\ref{fig20} and in the SCA in Fig.~\ref{fig21} shows that in both approximations the contribution from the EPI  is noticeably  smaller than those induced by the spin-fluctuation interaction. To explain this we note  a difference between the  self-energy for \,  the normal state  (\ref{29b}) and in the gap \, equation  (\ref{sc1}). While in summation over $\bf{q}$  contributions to the normal self-energy come from all symmetry components of interactions, in the gap equation   contributions are restricted only to the $B_{1g}$ symmetry component of  interactions determined by the symmetry of the $d$-wave gap $\phi_{\sigma}({\bf q})$. In particular,  a strong  momentum-independent EPI gives no contribution to the gap equation but results in a large contribution to the normal self-energy and the parameter  $Z(\bf{q}) $  in the gap equation  that suppresses $T_c$ (see also Ref.~\cite{Lichtenstein95}).  Therefore, the EPI  can be quite strong and gives observable polaronic effects but has a small $d$-wave partial harmonic and plays only a secondary role in the $d$-wave pairing. This results in the weak isotope effect on $T_c$ in the optimally doped cuprates.  The same holds for the intersite CI since only $d$-wave partial harmonic gives a contribution  to the gap equation.
\begin{figure}[htp]
\centering
\includegraphics[scale=0.45]{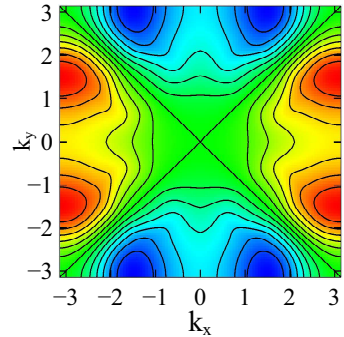}
\caption{(Color online) Wave-vector dependence of the superconducting gap at the FS $ \phi(\bf q)$.}
 \label{fig22}
\end{figure}
\begin{figure}[htp]
\centering
\includegraphics[scale=0.35]{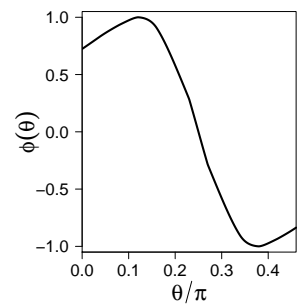}
\caption{Angle dependence of the superconducting gap $\phi(\theta)$ at the FS.}
 \label{fig23}
\end{figure}
The wave-vector dependence of the superconducting gap at the FS $ \phi(\bf q)$ at $\delta = 0.2$   is presented in Fig.~\ref{fig22} in the BZ. The angle dependence $ \phi(\theta)$  is shown in Fig.~\ref{fig23} where the angle $\theta$ is measured from the direction  $ M(\pi,\pi) \rightarrow X(\pi, 0)$  to the direction  $ M(\pi,\pi) \rightarrow Y(0,\pi)$.  We see that the maximum values of the gap are shifted in comparison with the model $d$-wave gap function $\phi^{(0)}({\bf q}) = \phi^{(0)} (\cos k_x -\cos k_y) $ from the BZ boundary at  $\,(0, \pm \pi), (\pm \pi, 0) \,$ points. Similar behavior was found in Ref.~\cite{Valkov03} in the $t$--$J^*$ model with  hopping parameters between  distant lattice sites.

\section{Summary}
\label{sec:6}

A detailed study  of the  electronic spectrum and superconductivity for strongly correlated electronic systems within the microscopic theory for the extended $t$--$J$--$V$ model~(\ref{1})  is presented. Besides the conventional AFM exchange interaction $J$, the EPI  and the intersite Coulomb repulsion are taken into account~(\ref{1a}). The projection technique  was employed to obtain the exact Dyson equation for the normal and anomalous (pair)  GFs~(\ref{15})
in terms of the Hubbard operators.  The self-energy (\ref{16}) given by many-particle GFs was calculated in the SCBA~(\ref{B5}), (\ref{B6}) in the second order of interactions.  The most important contribution is induced by the kinematical interaction for the  Hubbard operators~(\ref{2}). It results in a strong coupling of the electrons   with spin fluctuations of the order of hopping parameter $\, t({\bf q})$  much larger than the exchange interaction $J({\bf q})$.  Vertex corrections to this interaction in the SCBA should be small, of the order of  $\omega_s/\mu$, the spin-fluctuation energy $\omega_s$ to the Fermi energy $\mu$, as in the Eliashberg theory of EPI~\cite{Eliashberg60} with a small parameter $\omega_{ph}/\mu$ where $\omega_{ph}$ is a phonon energy.

In the generalized MFA, the first order of the projection technique, the electronic spectrum in Fig.~\ref{fig1} describes   well defined QP excitations. The FS in Fig.~\ref{fig2} shows a transformation from the four-pocket shape in the BZ at small doping to a large FS for higher doping. The superonducting $T_c$  induced by the exchange interaction $J$ in the MFA in  Fig.~\ref{fig4}  is large as in the RVB theory of Anderson~\cite{Anderson87} commonly used in many publications. However,  the intersite Coulomb repulsion strongly reduces the exchange interaction pairing and brings to  low $T_c$.

 The  self-energy contributions result in a strong renormalization both  the electronic spectrum and superconductivity.  The spectral density and the electron dispersion, Figs.~\ref{fig5} -- \ref{fig10},  reveal damped  excitations with the FS in the form of arcs at low doping, Figs.~\ref{fig14}, \ref{fig15}. The same strong coupling of  electrons   with spin fluctuations results in  high-$T_c$ for the $d$-wave pairing. The EPI and the intersite Coulomb repulsion   play a minor role for the $d$-wave pairing since they are determined  only by the $d$-wave partial harmonic in the gap equation. But the quasiparticle weight renormalization due to the normal state  self-energy  gives an order of magnitude smaller  $T_c$   in Fig.~\ref{fig21} in comparison with the  weak-coupling approximation in Fig.~\ref{fig20}.  The wave-vector dependence of the superconducting gap in Figs.~\ref{fig22}, \ref{fig23} clearly demonstrates the $d$-wave gap symmetry.

In general, the obtained results agree quite well with calculations for the Hubbard model and in a qualitative agreement  with  numerical calculations within  various cluster approximations, such  as CDMFT, CDA, CPT, and  with  ARPES experiments. In comparison with  the phenomenological spin-fermion models, we have no fitting parameter for  electron interaction with spin  fluctuations
which is given by two basic parameters of the model, the hopping parameter $t_{ij}$ and the AFM exchange interaction $J$. The HO technique permits to implement rigorously the constraint of no double occupancy  violated in  the MFA for the slave-fermion-(boson) theories.
 We believe that the spin-fluctuation pairing induced by the kinematical interaction may  be considered as  the mechanism of high-$T_c$ in cuprates.

\section{Appendix}
\label{sec:7}

To calculate the many-particle GFs in the self-energy  (\ref{29}),  (\ref{30})  we introduce the  time-dependent correlation function using the spectral representation, as, e.g.,
\begin{eqnarray}
 \langle\!\langle
B\sb{i\sigma\sigma'} X_l^{0\sigma' }
 | X_{l'}^{\sigma'0} B\sb{j\sigma\sigma'}^\dag
\rangle\!\rangle_{\omega}
= \frac{1}{2\pi}\int_{-\infty}^{\infty}
 dz \; \frac{e^{ z/ T}+1}{ \omega - z }
\nonumber \\
 \times\int_{-\infty}^{\infty}\! dt e^{i z t} \;\langle X_{l'}^{\sigma'0} B\sb{j\sigma\sigma'}^\dag
, B\sb{i\sigma\sigma'}(t)
X_l^{0\sigma' }(t)\rangle ,
 \label{a1}
\end{eqnarray}
In the SCBA (\ref{B5}) the many-particle time-dependent correlation functions are  presented as a product of single-particle correlation functions which are calculated self-consistently in terms of the corresponding GFs:
\begin{eqnarray}
\langle   X_{l'}^{\sigma 0} X_l^{0\sigma }(t)\rangle =
- \int_{-\infty}^{\infty} \frac{ d\omega' e^{-i\omega' t}}{\pi} n(\omega')
{\rm Im} G_{ll'}(\omega'),
 \nonumber\\
\langle B\sb{j\sigma\sigma'}^\dag
 |B\sb{i\sigma\sigma'}(t) \rangle
=- \int_{-\infty}^{\infty} \frac{ d\omega' e^{-i\omega' t}}{\pi}  N(\omega')
 {\rm Im} \langle\! \langle
B\sb{i\sigma\sigma'}|B\sb{j\sigma\sigma'}^\dag
 \rangle \! \rangle_{\omega'},
 \nonumber
\end{eqnarray}
where $n(\omega)$  and $N(\omega)$ are the Fermi and the Bose distribution functions, respectivly. Integration over time $t$ in Eq.~(\ref{a1})  yields for the spin-fluctuation contributions to the self-energy (\ref{29}):
\begin{eqnarray}
&& M^{sf}_{ij\sigma}(\omega) = \int \!\int_{-\infty}^{\infty} \frac{dz d\Omega}{ 2 \pi^2 \,Q}
\frac{\tanh(z/2T)+\coth(\Omega/2T) }{\omega -z - \Omega}
\nonumber\\
& &\times \sum_{l, l' \sigma'}
 t_{il}t_{j l'}\, {\rm Im} G_{l l'}(z)\,
{\rm Im} \langle\! \langle
B\sb{i\sigma\sigma'}|B\sb{j\sigma\sigma'}^\dag
 \rangle \! \rangle_{\Omega} .
   \label{a2}
\end{eqnarray}
Taking into account the definition of the $B_{i\sigma\sigma'} $-operator (\ref{14})
for the bosonic GFs we obtain the relation:
\begin{eqnarray}
&&\langle\! \langle B\sb{i\sigma\sigma'}|
B\sb{j\sigma\sigma'}^\dag
 \rangle \! \rangle_{\omega}
=  (1/4)\langle\! \langle N_{i} | N_{j}\rangle \!
\rangle_{\omega} \, \delta_{\sigma'\sigma}
 \nonumber \\
&&  + \langle\! \langle  S^z_{i}| S^z_{j}
 \rangle \!\rangle_{\omega} \, \delta_{\sigma'\sigma}
 + \langle\! \langle X^{\bar{\sigma}\sigma}_{i}|
 X^{\sigma \bar{\sigma}}_{j}\rangle \! \rangle_{\omega}
 \, \delta_{\sigma' \bar{\sigma}}.
 \label{a3}
\end{eqnarray}
After summation over  $\sigma'$ in (\ref{a2}) for the normal GF in the paramagnetic state, $
G_{ll'\sigma}(\omega) = G_{ll' \bar \sigma}(\omega)$, the spin-fluctuation contribution to the bosonic GF (\ref{a3}) takes  the form:
$\langle\! \langle  S^z_{i}| S^z_{j}
 \rangle \!\rangle_{\omega}
 + \langle\! \langle X^{\bar{\sigma}\sigma}_{i}|
 X^{\sigma \bar{\sigma}}_{j}\rangle \! \rangle_{\omega} =
 \langle\! \langle {\bf S}_{i}|{\bf S}_{j}
 \rangle \!\rangle_{\omega} $.
 Introducing the Fourier representation similar to (\ref{8b}) we obtain for the spin-fluctuation contribution to the self-energy
 \begin{eqnarray}
M^{sf}({\bf  k},\omega) =\int_{-\infty}^{+\infty} \!\!\! \int_{-\infty}^{+\infty} \frac{dz d\Omega}{2\pi^2 \,Q}\, \frac{\tanh(z/2T)+\coth(\Omega/2T) }{\omega-z-\Omega}
 \nonumber \\
\frac{1}{ N} \sum_{\bf  q} |g_{sf}({\bf k,  q})|^{2} {\rm Im} \chi\sb{sf}({\bf k- q},\Omega)\mbox{Im} G({\bf q},z).
\label{a4}
\end{eqnarray}
Calculations for the CI and the EPI  give the corresponding contribution to the self-energy (\ref{29}):
\begin{eqnarray}
 M^{c,\, ep}_{ij\sigma}(\omega) = \int \!\int_{-\infty}^{\infty}  \frac{dz d\Omega}{2\pi^2\,Q}
\frac{\tanh(z/2T)+\coth(\Omega/2T) } {\omega -z - \Omega}
\nonumber\\
\times \Big\{  \sum_{l l'}
 V(i,l)  V^*(j,l')  {\rm Im} G_{ij}(z)\,
{\rm Im} \langle\! \langle
N_{l}|N_{l'}^\dag \rangle \! \rangle_{\Omega}
  \nonumber\\
+\sum_{l l'}\,g_{i l} \,g^*_{j l'}
 {\rm Im} G_{ij}(z)\,
{\rm Im} \langle\! \langle
u_{l}|u^\dag_{l'} \rangle \! \rangle_{\Omega} \Big\}.
  \label{a5}
\end{eqnarray}
As a result, after  the Fourier transformation of all the contributions in equation (\ref{29})
we obtain the normal self-energy given by Eq.~(\ref{29a}).

Similar calculations for the anomalous component of the self-energy result in the equation
\begin{eqnarray}
 \Phi_{ij \sigma}(\omega)
& = &\frac{1}{Q} \sum_{l,l'} \Big\{\sum_{\sigma'}\, t_{il} t_{j l'}
\langle\langle B_{i \sigma \sigma'} X_{l}^{0\sigma'} |  B_{j \bar{\sigma} \bar{\sigma}'} X_{l'}^{0\bar{\sigma}'} \rangle\rangle_\omega
\nonumber\\
& - &  V(i,l)  V(j,l')
\langle\langle X_{i}^{0\sigma}\, N_{l} |  X_{j}^{0 \bar{\sigma}} N_{l'} \rangle\rangle
\nonumber\\
& - &\,g_{i l} \,g_{j l'}  \langle\langle X_{i}^{0\sigma} u_{l} |
 X_{j}^{0 \bar{\sigma}}
 u_{l'} \rangle\rangle_\omega\Big\} .
 \label{a6}
\end{eqnarray}
For the many-particle anomalous GF  we have
\begin{eqnarray}
 \langle\langle B\sb{i\sigma\sigma'} X_{l}^{0 \sigma'}
 | B\sb{j\bar{\sigma}\bar{\sigma}'} X_{l'}^{ 0 \bar{\sigma}'}
\rangle\!\rangle_{\omega}
 =  \frac{1}{2\pi}\int_{-\infty}^{\infty}
 dz \frac{e^{ z / T}+1}{\omega - z}
 \nonumber\\
\times\int_{-\infty}^{\infty} dt e^{izt}\, \langle  X_{l'}^{0 \bar{\sigma}' } B\sb{j\bar{\sigma}\bar{\sigma}'}
 | B\sb{i\sigma \sigma'}(t) X_{l}^{0 \sigma' }(t)\rangle .
  \label{a7}
\end{eqnarray}
Then  using the SCBA (\ref{B6}) we calculate
the time-dependent correlation functions self-consistently using the corresponding anomalous GFs.
Integration over time $t$ in Eq.~(\ref{a7})  yields for the  anomalous self-energy:
\begin{eqnarray}
  && \Phi _{ij\sigma}(\omega) =
 \int \!\int_{-\infty}^{\infty} \frac{dz d\Omega}{ 2 \pi^2 \,Q} \;  \frac{\tanh(z/2T)+\coth(\Omega/2T)}
  {\omega -z  - \Omega}
\nonumber\\
&& \times  \sum_{l,l'} \Big\{\sum_{\sigma'}\, t_{il} t_{j l'}
 \,{\rm Im}\langle\! \langle  X_{l}^{0\sigma'}
  | X_{l'}^{0 \bar{\sigma}'}  \rangle \! \rangle_{z} \,
 {\rm Im} \langle\! \langle
B\sb{i\sigma\sigma'}| B\sb{j\bar{\sigma}\bar{\sigma}'}
 \rangle \! \rangle_{\Omega}
  \nonumber\\
& - &  V(i,l)  V(j,l')
{\rm Im}\langle\! \langle  X_{i}^{0\sigma}
  | X_{j}^{0 \bar{\sigma }}  \rangle \! \rangle_{z} \,
 {\rm Im} \langle\langle  N_{l} | N_{l'} \rangle\rangle_\Omega
\nonumber\\
& - &  {\rm Im}\langle\! \langle  X_{i}^{0\sigma}
  | X_{j}^{0 \bar{\sigma }}  \rangle \! \rangle_{z} \,
 {\rm Im}\langle\langle  u_{l} | u_{l'} \rangle\rangle_\Omega \Big\}.
  \label{a8}
\end{eqnarray}
Here for  the bosonic GF we have  the relation
\begin{eqnarray}
\langle\! \langle B\sb{i\sigma\sigma'}|
 B\sb{j\bar{\sigma}\bar{\sigma}'}
 \rangle \! \rangle_{\omega}
=  (1/4)\langle\! \langle N_{i} | N_{j}\rangle \!
\rangle_{\omega} \, \delta_{\sigma'\sigma}
 \nonumber \\
  - \langle\! \langle  S^z_{i}| S^z_{j}
 \rangle \!\rangle_{\omega} \, \delta_{\sigma'\sigma}
 + \langle\! \langle X^{\bar{\sigma}\sigma}_{i}|
 X^{\sigma \bar{\sigma}}_{j}\rangle \! \rangle_{\omega}
 \, \delta_{\sigma' \bar{\sigma}}.
  \label{a9}
\end{eqnarray}
Summation over $\sigma'$  for the bosonic GF (\ref{a9}) and the anomalous GF
 $ F_{ll'\sigma}(\omega) = - F_{ll' \bar \sigma}(\omega)$  in Eq.~(\ref{a8})
results in the relation: $\, - \langle\! \langle S^z_{i}| S^z_{j}
\rangle\!\rangle_{\omega}\,
 F_{ll'\sigma}(\omega)
  + \langle\! \langle X^{\bar{\sigma}\sigma}_{i}|
 X^{\sigma \bar{\sigma}}_{j}\rangle \! \rangle_{\omega} \,
 F_{ll' \bar \sigma}(\omega)  = -
 \langle\! \langle {\bf S}_{i}|{\bf S}_{j}
 \rangle \!\rangle_{\omega}\, F_{ll'\sigma}(\omega) $.
After  the Fourier transformation of all the contributions in (\ref{a8})
we obtain the anomalous self-energy  component  (\ref{30a}).\\

{\bf Acknowledgement}\\

We would like to thank V.S. Oudovenko for helpful discussions.

\end{document}